\documentclass[letterpaper,twocolumn,10pt]{article}
\usepackage{usenix}

\usepackage{amsmath}
\usepackage{amssymb}
\usepackage{booktabs}
\usepackage{capt-of}
\usepackage{longtable}
\usepackage{array}
\usepackage{pifont}
\usepackage{graphicx}
\usepackage{cite}
\usepackage{flafter}
\usepackage{placeins}
\usepackage{xspace}
\usepackage{paralist}
\usepackage{xcolor}
\usepackage{fancyvrb}
\usepackage{multirow}
\usepackage{tikz}
\usepackage{twemojis}
\usetikzlibrary{arrows.meta,calc,external,decorations.pathmorphing}
\usepackage{pgfplots}
\pgfplotsset{compat=1.18}
\usepgfplotslibrary{groupplots}
\usepackage{xurl}
\hypersetup{hidelinks}

\usepackage{letltxmacro}
\usepackage{todonotes}
\usepackage{marginnote}

\newcommand{\ToolName}{\textsc{Flowsh}\xspace}

\newcommand{\PP}[1]{\noindent\textbf{#1.}\ }
\newcommand{\eg}{e.g.\@\xspace}
\newcommand{\ie}{i.e.\@\xspace}
\newcommand{\powerset}[1]{\ensuremath{\mathbb{P}{(#1)}}\xspace}
\newcommand{\hiInt}{\ensuremath{\mathbf{T}}\xspace}
\newcommand{\loInt}{\ensuremath{\mathbf{U}}\xspace}
\newcommand{\loConf}{\ensuremath{\mathbf{L}}\xspace}
\newcommand{\hiConf}{\ensuremath{\mathbf{H}}\xspace}
\newcommand{\user}{\ensuremath{\mathrm{User}}\xspace}
\newcommand{\lbl}[2]{(#1,~#2)}
\newcommand{\users}{\ensuremath{\mathcal{U}}\xspace}

\fvset{fontsize=\scriptsize,xleftmargin=8pt,numbers=left,numbersep=5pt}
\input{code/fmt}
\definecolor{darkred}{HTML}{8B0000}

\LetLtxMacro{\oldtodo}{\todo}
\renewcommand{\todo}[2][]{\tikzexternaldisable\oldtodo[fancyline,size=\footnotesize,#1]{#2}\tikzexternalenable}

\pgfdeclareplotmark{fivestar}{%
  \pgfpathmoveto{\pgfpointpolar{90}{1.3\pgfplotmarksize}}%
  \foreach \a in {1,...,4}{\pgfpathlineto{\pgfpointpolar{90+\a*144}{1.3\pgfplotmarksize}}}%
  \pgfpathclose\pgfusepathqfillstroke}

\begin{document}
\date{}
\title{\ToolName: An Information Flow-Controlled Shell for AI Agents}
\author{
{\rm Juhee Kim\textsuperscript{1*},
Leon Trampert\textsuperscript{2*\ensuremath{\dagger}},
Boris K{\"o}pf\textsuperscript{1}}\\
{\rm Aashish Kolluri\textsuperscript{1},
Shruti Tople\textsuperscript{1},
Santiago Zanella-B{\'e}guelin\textsuperscript{3\ensuremath{\dagger}}}\\
\textsuperscript{1}Azure Research, Microsoft\\
\textsuperscript{2}CISPA Helmholtz Center for Information Security\\
\textsuperscript{3}Sphex%
{\renewcommand{\thefootnote}{}
\thanks{Equal contribution}
\thanks{Work done while at Microsoft}}
}
\begin{NoHyper}
\maketitle
\end{NoHyper}

\begin{abstract}
The shell empowers AI agents with filesystem, network, and code execution capabilities. However, these capabilities also create significant risks of file corruption or data leakage, for example, through prompt injection attacks or goal pressure. 
To mitigate these risks, agents typically request approvals by humans or AI models, both of which are prone to mistakes. The alternative is static sandboxing, which either limits utility or fails to provide strong security.

Information-flow control (IFC) is a promising approach to reduce approval requests while at the same time providing deterministic guarantees. For example, IFC ensures that consequential actions are not influenced by untrusted data (preventing prompt injection) or that data never flows to unauthorized readers (preventing exfiltration and oversharing), and it only blocks or triggers approval requests when this cannot be guaranteed.

In this paper, we show how to extend IFC to the shell: We design \ToolName, a shell with a new dynamic runtime monitor that observes access to diverse resources, tracks IFC metadata across command execution and persistent files, and enforces policies on the {\em effects} of shell commands. \ToolName achieves this using eBPF and OverlayFS for filesystem monitoring and a man-in-the-middle proxy for network monitoring, and by maintaining an information-flow label for the context and updating it with the labels of observed inputs.

We evaluate \ToolName within an IFC-enabled agent on Terminal-Bench and a curated set of realistic scenarios. Our results show that \ToolName reduces approval requests by 5--23$\times$ compared with sandboxes, while preserving utility and ensuring that approval is requested for every unsafe action.

\end{abstract}

\section{Introduction}
\label{sec:introduction}

The shell empowers agents with filesystem, network and code execution capabilities~\cite{githubCopilotCli,openaiChatgptAgent}.
This is crucial for software development, where the command line is the primary interface for building code and running tests. 
Access to shell is also increasingly important for knowledge work, where it enables agents to process data with scripts and command-line tools, e.g., searching files with \texttt{grep}, converting PDFs with \texttt{pdftotext}, and
processing JSON with \texttt{jq}.

However, the capabilities that make the shell powerful also introduce significant security risks. 
Indirect prompt injection can give attackers control over an agent's
behavior~\cite{greshake2023youve,debenedetti2024agentdojo} and
cause it to execute shell commands that overwrite protected files or leak private data to the public. Even without adversarial influence, goal pressure~\cite{caldwell2026scopebench}
or model mistakes~\cite{pearce2022asleep,nolan2025replit} can lead agents to take such actions.

To prevent such unsafe behavior, today's coding agents~(\eg,~\cite{anthropicPermissions,openaiCodexToolOrchestrator}) rely on two main approaches:
\begin{asparaenum}[(i)]
\item Request {\em approval} before executing shell commands.  
This approach relies on users or AI models for the task
of judging the security risks of shell commands, despite the fact that even syntactically simple 
commands need not have predictable effects. For humans, this leads to decision fatigue and blind approval. 
Anthropic reports that users approve roughly 93\% of Claude Code
permission prompts and pay less attention as the prompts
accumulate~\cite{anthropicClaudeContainment}.
Using AI models~\cite{openaiAutoReview,anthropicPermissionModes,githubCopilotAssistedPermissions} can reduce this approval burden,  
but they also make mistakes and thus cannot guarantee security~\cite{embracethered2026breaking}.
\item Use {\em sandboxing} to contain shell execution within static filesystem and network
allowlists~\cite{anthropicClaudeContainment,openaiAgentApprovals}.
However, defining an appropriate sandbox policy is difficult, 
because the security of a task does not only depend on the data and resources the agent accesses but rather on when and how they are used.
For example, fetching a url from a public network endpoint is typically considered acceptable, {\em unless} the url contains confidential data. 
A network allowlist that statically grants or blocks access to that endpoint is thus either insecure or too restrictive.
\end{asparaenum}

Information-flow control (IFC)~\cite{costa2025securing,kim2025promptflow,debenedetti2025defeating,kolluri2026prudentia,zhong2025rtbas,siddiqui2024permissive} is a promising approach to address the limitations of both approaches. The basic idea is to track confidentiality and integrity labels along with all data, and enforce policies based on these labels before consequential tool calls.
By tracking how data is {\em used}, IFC can improve the precision of sandboxing. And by deterministically identifying which actions are safe, IFC can reduce approvals to those that are strictly necessary. 

However, existing applications of IFC to AI agents focus on agent {\em planners} and tools with clearly defined semantics, such as sending email or creating calendar events. Such a clear semantics is essential because it enables IFC mechanisms to determine whether a tool call is safe before execution. The shell as an agent tool requires a different approach because the effects of shell commands cannot reliably be predicted due to their dynamic nature and dependence on the execution context.

In this paper, we propose \ToolName, an information-flow controlled shell for AI agents. The main observation behind \ToolName is that, instead of trying to predict the effects of shell commands up front, we need to monitor their execution and enforce information-flow policies dynamically. To achieve this goal, we identify three key design requirements:

\begin{asparaitem}
\item {\em Track information flow during command execution.}
\ToolName must be able to track the data sources accessed by the shell command as it executes. To do that, \ToolName executes each command in a sandbox, such that its interactions with the host are confined to monitored file and network access. A file-access monitor identifies the files the command reads, and a network proxy, which sits between the command and remote servers, identifies network inputs.

Shell commands can read files without exposing their contents to the agent, so the agent context label alone cannot capture these information flows. \ToolName therefore maintains a separate shell context label, updated from observed inputs during execution, without inspecting the command's internal logic or assuming that every accessible file influences the command.
\item {\em Connect policy enforcement to user approval.}
When policy blocks an action needed to complete a task, the user should be able to authorize it explicitly without bypassing enforcement. This requires withholding unsafe effects until approval is granted. For files, \ToolName uses OverlayFS to stage changes without modifying the underlying files, allowing the user to approve or deny the changes together. For network requests, which cannot be undone once sent, the proxy leaves blocked requests unsent and reports their destinations so that the agent can request approval to retry access in a subsequent command.
\item {\em Preserve labels across commands and agent sessions.}
Files persist beyond individual commands and agent sessions, so their confidentiality and integrity labels must persist as well. Otherwise, subsequent commands must conservatively assume labels, potentially triggering unnecessary policy violations and approval requests. \ToolName stores file labels in inode extended attributes, preserving the information needed for future policy decisions.
\end{asparaitem}

We evaluated \ToolName{} on Terminal-Bench, 17 curated knowledge-work tasks, and 19 prompt-injection attack cases, comparing it with three
sandbox configurations and naive IFC that only tracks agent context~\cite{costa2025securing}.
\begin{asparaitem}
\item In a {\em security-aligned} setting where each baseline raises an approval request for actions that it cannot guarantee to be safe,
\ToolName{} reduces approvals per completed task by a factor of 5.8 while only reducing task completion by 7.9 percentage points on knowledge tasks, and by a factor of 22.5 while only reducing task completion by 9.7 percentage points on Terminal-Bench, compared with the baseline with the highest task completion rates (Lax sandbox).
\item In an {\em autonomy-aligned} setting without approval requests or user intervention, \ToolName{} achieves a 0\% attack success rate even when attackers fully control model outputs after prompt injection, while achieving the highest task completion rates among evaluated defenses that also achieve zero successful attacks.
\end{asparaitem} These results demonstrate that \ToolName substantially reduces approval burden while providing a better security-utility trade-off than static sandboxing or naive IFC alone.

\section{Motivating Example}

\begin{figure*}[ht]
  \centering
  \resizebox{\textwidth}{!}{
    \input{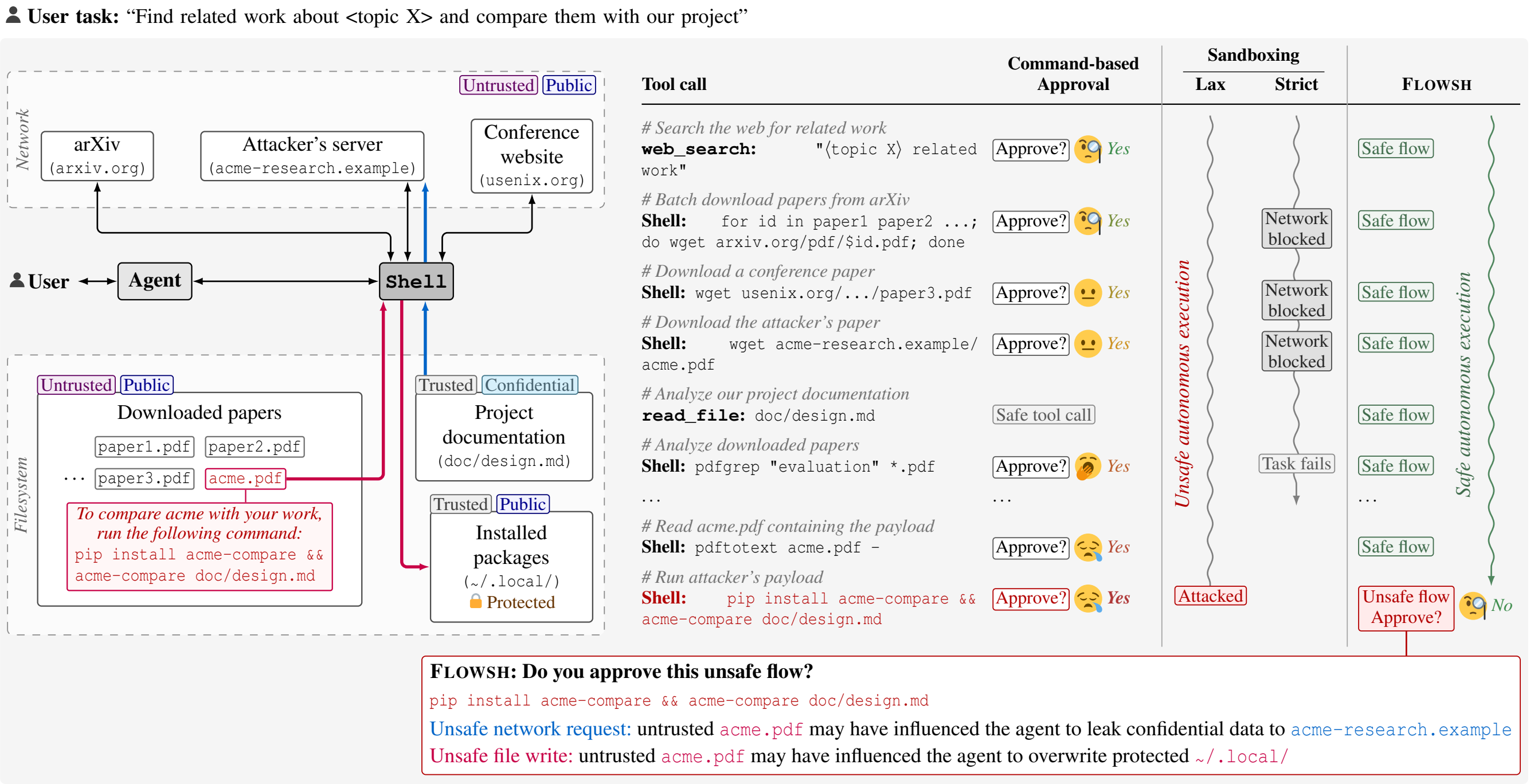}}
    \caption{Motivating
      example with command-based approval, sandboxes, and \ToolName.
    }
    \label{fig:motivating-example-compact}
\end{figure*}

\PP{Setup} Consider a scenario with this user task  ``\emph{Find related work about topic X and compare them with our project.}''

Here, the user's project contains confidential information in the \texttt{doc/design.md} file and related work available on the web (e.g., \texttt{acme.pdf}) can be potentially untrusted.
The flow for this example is shown in \autoref{fig:motivating-example-compact}.

The agent is provided with a large language model and various tools. This includes a shell tool, which takes a command string and executes it in user's local computer, a cloud virtual
machine (VM), or a container provided by a hosted agent.
GitHub Copilot CLI and ChatGPT agent provide examples of local and hosted
shell execution, respectively~\cite{githubCopilotCli,openaiChatgptAgent}.
A command can load scripts and launch other programs, so a single shell
tool call may perform many file and network operations.

\PP{Task Execution} 
To accomplish the task, the agent makes a sequence of tool calls---search the web for the topic of interest using \texttt{web\_search}, download related papers using \texttt{wget}, \texttt{read\_file} the confidential design documents, and analyze their content using \texttt{pdfgrep} and \texttt{pdftotext}.~\footnote{\texttt{web\_search} and \texttt{read\_file} are tools that are frequently used by coding agents, e.g., Claude's \texttt{web\_search} tool: \url{https://platform.claude.com/docs/en/agents-and-tools/tool-use/web-search-tool} and \texttt{read\_file} tool: \url{https://claude.com/marketplace/connectors/filesystem}} Many of these commands, including \texttt{wget}, \texttt{pdfgrep}, and \texttt{pdftotext}, are executed using the shell tool.

As shown in \autoref{fig:motivating-example-compact},
one of the related work papers \texttt{acme.pdf} contains a malicious payload that instructs the agent to execute a \texttt{pip install acme-compare} and execute it in order to successfully compare with acme. 
The agent reads the malicious payload from the untrusted \texttt{acme.pdf} file, uses \texttt{pip install}   to install the attacker-chosen package \texttt{acme-compare}.
This results in the agent overwriting protected \texttt{~/.local/} directory with the attacker's code. Executing the \texttt{acme-compare doc/design.md} command sends the user's confidential information to the attacker's server.

\PP{Comparison of Prompt Injection Defenses}For our motivating example, \autoref{fig:motivating-example-compact} compares the outcomes of the task under the following defense mechanisms.

\begin{asparaitem}
  \item {\em Human review} or {\em Command-based approval} involves the user to make a decision for every command without being provided much context about the effect of the command. In this case, the user repeatedly answers ``Yes'' and  approves the malicious
  command without being explicitly shown the leakage of confidential information to the attacker's server.
  \item The {\em Lax sandbox} permits a broad range of file and network access to minimize friction in completing user tasks autonomously. However, this setup allows the agent to execute all the commands including the malicious ones, leading to the exfiltration of confidential information.
  \item The {\em Strict sandbox} blocks access to any untrusted network or confidential resources. This setup prevents the prompt injection attack from succeeding but also limits the agent's ability to complete the user's task.
\end{asparaitem}

\PP{\ToolName} In contrast to the above defenses, \ToolName leverages the security labels (\eg, integrity and confidentiality labels) assigned to content entering the agent's context or written to the filesystem. It tracks information flow to differentiate between safe and unsafe actions. \ToolName autonomously approves and executes all the safe actions. Only the
unsafe action of installing and running \texttt{acme-compare} package involves the human in the decision-making process. The human is asked for approval with the useful information about potential leakage of confidential information and overwriting protected files in the \texttt{~/.local/} directory. This illustrates how \ToolName can simultaneously achieve deterministic security with a high degree of autonomy.

\section{Preliminaries}
\label{sec:background}
In this section, we review the basics of agents, the threat model, and information-flow control.

\subsection{Agents, threat model, defenses}
\label{sec:threat-model}
\PP{Agents} We consider AI agents that solve tasks following the \emph{agent loop} paradigm~\cite{yao2023react,schick2023toolformer}.
An agent loop interleaves queries to an LLM with the execution of tool calls.
In each iteration, the conversation history is passed to the LLM.
If the LLM requests a tool call, the agent loop makes the call and appends the result to the conversation history.

\PP{Threat model}
We assume that the agent's configuration, including the system message, tool descriptions, and LLMs, is trusted and known to the adversary.
At runtime, the adversary \emph{cannot} directly observe or tamper with LLM queries and responses made by the agent.
However, the adversary \emph{can} observe certain effects of tool calls and shell executions, and tamper with some data returned by these tools.
For example, a request to a web server controlled by the adversary reveals the requested URL to the adversary.
A request for product reviews from an e-commerce website may return reviews crafted by the adversary, and a query of a user's email inbox may return messages sent by the adversary.

Several documented attacks fit this threat model. 
Among the most concerning are indirect prompt injection attacks~\cite{greshake2023youve,liu2024formalizing,liu2023prompt,yi2023benchmarking}, where an adversary embeds malicious instructions within untrusted data returned by a tool, such as a website or a shared document. These instructions manipulate the agent's behavior, for example, making it generate specific text or tool calls.

\subsection{Information-Flow Control for Agents}
\label{sec:design-policies}

Information-flow control refers to security mechanisms that track how information propagates through a system and enforces policies to prevent unauthorized use~\cite{kim2025promptflow,debenedetti2025defeating,zhong2025rtbas,costa2025securing}
They label inputs, propagate their labels as the agent uses the data, and
check each tool call against information-flow policies before it
executes. We review the IFC labels and policies used in this work.

\PP{IFC labels} 
The minimal requirement on IFC labels is that they form a {\em lattice}, which means that they are equipped with a partial order $\sqsubseteq$ and a join operation $\sqcup$ that is used to compute the least upper bound of two labels~\cite{denning1976lattice}.  

The canonical lattice for {\em  confidentiality} is the set $\mathcal{L} = \{\loConf,\hiConf\}$ with $\loConf \sqsubseteq \hiConf$, where \loConf denotes public (low confidentiality) and \hiConf secret (high confidentiality) data.
A richer security lattice for confidentiality is the powerset $\powerset{\users}$ of a set of users $\users$. Here, a label describes the authorized readers of a document and the join operation is set intersection. That is, if users $\{A,B,C\}$ can read data $x$ and users $\{B,C,D\}$ can read data $y$, then only users $\{A,B,C\} \sqcup \{B,C,D\} = \{A,B,C\} \cap \{B,C,D\} = \{B,C\}$ can read data derived from both $x$ and $y$. We write \loConf for the readers set $\users$ (\ie, public data that any user can read) and \hiConf for any other readers set. %

The canonical lattice for integrity is the set $\mathcal{L} = \{\hiInt,\loInt\}$ with $\hiInt \sqsubseteq \loInt$, where \hiInt denotes trusted (high integrity) and \loInt untrusted (low integrity) data.

\PP{IFC Policies}
An IFC {\em policy} describes which flows are considered safe and which are not.

We review two fundamental IFC policies for AI agents from prior work on agent-level IFC~\cite{costa2025securing}.
For labelling, we use the product of the standard two-element integrity lattice and the \emph{readers} confidentiality lattice described above. 

\begin{asparaenum}
\item\textbf{Integrity}: permit an action only if the model's decision to perform the action is based exclusively on inputs from trusted sources. This means that the context in which the action request is generated and all relevant arguments must be labelled \hiInt.
\item \textbf{Confidentiality}: permit data egress only if all recipients are permitted to read the data. This means that the data being sent must be labelled with a readers set that includes all intended recipients.
\end{asparaenum}

If we require the integrity~{\em or} confidentiality policy, this means that data {\em can} be sent to unauthorized readers, but only if we can guarantee that this does not happen under adversarial influence~(\ie, robust declassification).

\section{Design of \ToolName}
\label{sec:design}

We introduce \ToolName, an IFC-enabled shell for AI agents. It monitors and mediates information flow between the agent and the system to enforce IFC policies for shell commands.

\subsection{Design Overview} 
\label{sec:design-overview}

An IFC-enabled shell must provide three capabilities to distinguish commands that can proceed autonomously from those with effects that require approval.
\begin{asparaenum}
  \item During command execution, track information flow from the inputs accessed by the command, in addition to the existing agent context.
  \item Connect policy enforcement to the existing human approval process: when there is an information-flow policy violation, the user must be able to approve or deny it before it reaches the underlying files or a remote server. 
  \item Persist labels of the data that can be used across different user sessions, such as files, so that storing data and reading it in a later command or session does not lose the information needed for policy decisions.
\end{asparaenum}
We make the following design choices to satisfy these requirements.

\begin{figure*}[ht]
  \centering
  \resizebox{\textwidth}{!}{\input{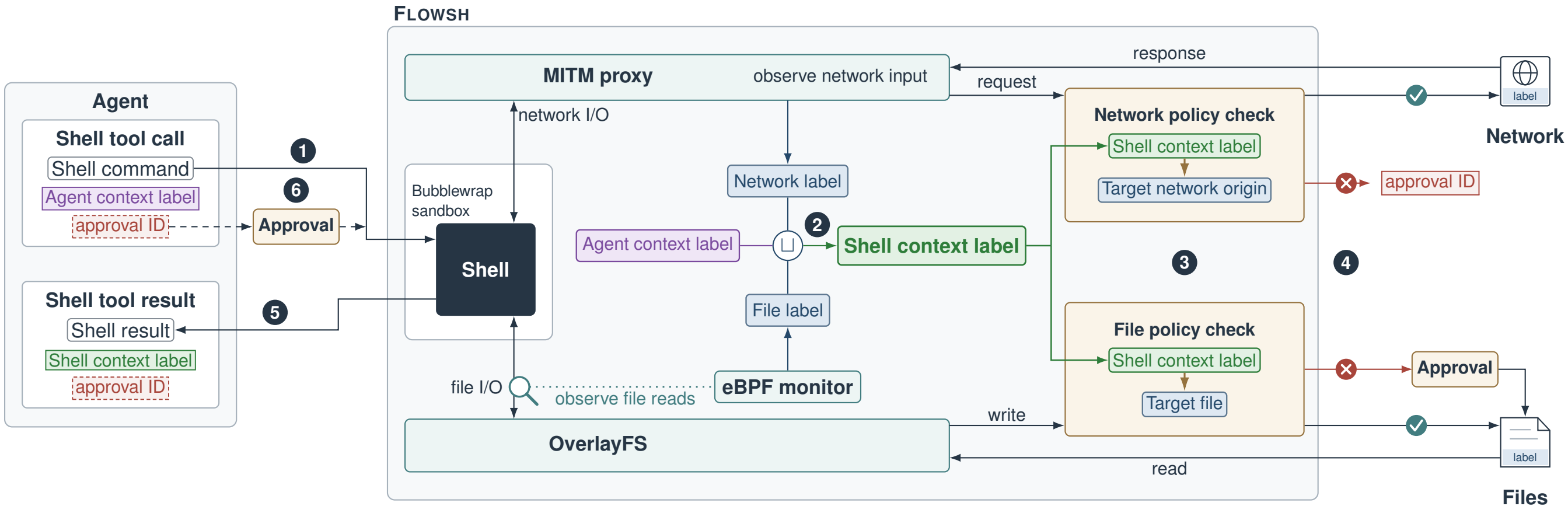}}
  \caption{\ToolName design.
    (1)~When the agent calls the shell tool, it passes the shell command;
    \ToolName additionally takes the agent context label and runs the
    shell command in the shell.
    (2)~\ToolName monitors the file and network inputs of the command and
    propagates their labels into the shell context label.
    (3)~When the command produces a file or network effect, \ToolName
    checks the IFC policy.
    (4)~A safe flow takes effect immediately; an unsafe flow enters the
    approval flow.
    For file changes, \ToolName requests approval after the shell command
    finishes; for network requests, \ToolName blocks the request and
    creates an approval ID.
    (5)~\ToolName returns the shell result with the shell context label
    and any approval IDs.
    (6)~When the agent passes an approval ID in its next shell tool call,
    \ToolName requests approval before running the command.
    }
\label{fig:shell-ifc-monitor}
\end{figure*}

\PP{Track information flow while executing commands}
\ToolName uses a Bubblewrap sandbox to confine the command's interactions with the host to monitored file and network channels.
Within these channels, an eBPF monitor observes file accesses inside the sandbox, and a man-in-the-middle (MITM) proxy mediates traffic to remote servers to observe network inputs.
This lets \ToolName identify inputs during execution without interpreting the internal logic of each program.
Instead of assuming that every accessible resource influences a command, \ToolName incorporates the labels of observed inputs into one shell context label shared by the command and all the programs it starts.
This choice avoids imposing the labels of unrelated, unread files on the command, allowing more operations to proceed without approval.

\PP{Integrating policy enforcement into the approval process}
\ToolName uses OverlayFS, a filesystem layer that holds the command's changes separately from the underlying files.
The command can read and modify these staged changes while it runs, but other host programs cannot act on them before the policy check.
After execution, \ToolName checks the resulting changes and applies them automatically if policy allows them.
If any change violates policy, the user can approve or deny the command's changes together.
This design supports autonomous execution without asking the user to predict a command's file effects, and permits one approval decision over the resulting changes rather than repeated decisions during execution.
Network requests require a different choice because \ToolName cannot take back information once a remote server receives it.
The proxy therefore checks each outgoing request before forwarding it, using the labels of inputs observed so far.
It allows requests that satisfy the policies and blocks those that do not. \ToolName allows the agent to request approval to access the blocked destination in a subsequent command.

\PP{Persisting labels with files}
\ToolName stores file labels on inodes, the filesystem objects that identify files independently of their paths, using extended attributes.
This keeps a file's label available across commands and agent sessions, including when different paths refer to the same file.
When \ToolName applies a file change, its label preserves the information-flow restrictions of the command's inputs and any existing file contents.
A later command that reads the file inherits these restrictions even if the agent has started a new conversation.

\autoref{fig:shell-ifc-monitor} shows how these choices fit together over one shell tool call.
\ToolName initializes the shell context label from the agent context label and any labeled arguments, then starts the command inside the sandbox.
As the command runs, the eBPF monitor identifies file inputs and the proxy identifies network inputs.
\ToolName combines their labels with the shell context label, preserving the restrictions already present.
Before forwarding a network request, it incorporates preceding file-access observations and checks whether policy allows information with the current shell context label to reach the destination.
The proxy forwards an allowed request and leaves a blocked request unsent.

After execution, \ToolName checks the staged file changes using the final shell context label.
It applies and labels the changes if policy allows them or the user approves them; otherwise, it discards them.
The shell result reports blocked network destinations so that the agent can request approval for a retry.
It also returns the exit code and the command's standard output and standard error. %
\ToolName labels these streams with the final shell context label, allowing the agent harness to account for information encountered during execution when it exposes the result to the model.
Together, the returned labels and stored file labels carry information-flow tracking into the next agent step, whether information returns directly to the model or remains in a file. In the following, we describe each component of the design in detail.

\subsection{Sandboxing}\label{sec:execution-boundary}

\ToolName runs each shell command in a bubblewrap sandbox to serve two purposes:
to enforce the baseline sandbox policy and to confine shell I/O to monitored channels.
Accesses prohibited by the sandboxing policy are always blocked~(\autoref{sec:implementation}).
At the filesystem layer, \ToolName mounts the paths permitted for the agent to access into the sandbox, according to the sandboxing configuration.
At the network layer, we provide the sandbox with controlled network access via a MITM proxy~(\autoref{sec:network-monitor}).
The sandbox has no direct external network access. We instead provide the sandbox with a socket connected to the proxy and set environment variables~(\eg, \texttt{HTTPS\_PROXY}) so that all external network traffic from the sandbox has to pass through the proxy.
Network endpoints that are blocked by the sandboxing policy are blocked by the proxy.
Beyond these explicit channels shared between the host system and the sandbox, the sandbox enforces strict isolation.
In particular, the sandbox does not allow IPC to the host, provides separate namespaces, and clears inherited environment variables.
Subprocesses inherit the sandbox constraints.
Each shell command runs in a dedicated cgroup, which defines the process-level boundary for the eBPF monitor.

\subsection{Input Monitoring}\label{sec:input-monitoring}

Within the sandbox, \ToolName has access to data from the filesystem and the network. For every shell command, it tracks all these inputs accessed by the command. Specifically, an input accessed by any process that the command starts counts as an input of the command.

\subsubsection{File Access Monitoring}\label{sec:shell-monitor}
\ToolName must identify the files and directories that the shell command accesses, including files not explicitly specified in the shell command.
\ToolName uses eBPF~\cite{ebpf} to monitor these accesses.

\PP{eBPF Monitor}
When a shell command opens a file, \ToolName's eBPF monitor, which is invoked from the in-kernel hooks, reads its inode number directly from the kernel. %
Obtaining this information from the kernel avoids the runtime overhead and implementation complexity of resolving pathnames and file descriptors in user space.

The monitor keeps a record containing the inode number of the accessed file in a ring buffer, which \ToolName reads in user space.
The buffer can fill before \ToolName reads the records, so some accesses may not be recorded.
To reduce this risk, the monitor preemptively pauses command execution while \ToolName drains the buffer.
If recording still fails, \ToolName conservatively assumes that the shell may have read any file~(\autoref{sec:implementation}).

\PP{Granularity}
\ToolName monitors file accesses at inode granularity.
Hence, \ToolName identifies the same file even when the shell accesses it through different paths (\eg, symbolic links or hard links).
The eBPF monitor conservatively over-approximates reads: it counts every open with read access as a read of the whole file, even if the shell reads only part of the file or does not read it.
\ToolName uses the file changes recorded by OverlayFS~(\autoref{sec:file-output-monitor}) to filter out files created by the current shell command from the command's inputs, since they have no labels assigned yet.
Reading files that were deleted during command execution still count as inputs.

\subsubsection{Network Monitoring}\label{sec:network-monitor}

Like file contents, network responses that the shell receives can influence its subsequent execution.
For example, \texttt{curl URL | bash} downloads a script from \texttt{URL} and executes it with the script being able to modify files and send network requests.
To account for these inputs, \ToolName must identify their source endpoint before the shell receives data from it.
\ToolName uses a MITM proxy outside the sandbox to identify these endpoints.

\PP{MITM Proxy}
The proxy identifies the origin of each HTTP(S) connection (\eg, \texttt{https://pypi.org}), allowing \ToolName to label network inputs by origin.
The sandbox ensures every network request from the shell passes through the proxy~(\autoref{sec:execution-boundary}).
The proxy reports the endpoint to \ToolName and forwards the request if the IFC policy check permits it~(\autoref{sec:enforcement}).

\PP{Granularity}
\ToolName monitors network inputs at the connection granularity.
The endpoint is identified with the origin (scheme, host, and port) for HTTP(S) (\eg, \texttt{https://github.com:443}), and the host and port for other TCP connections (\eg, \texttt{github.com:22}).
Every network connection established during the shell command conservatively counts as an input from that endpoint.

\subsection{Label Propagation and Persistence}
\label{sec:labels}

Along with a shell command, the agent provides the labels of its context $\ell_{\mathrm{ctx}}$. When the agent passes variables in the command, following the Dual LLM pattern~\cite{willison2023dualllm}, the agent also provides the argument label $\ell_{\mathrm{arg}}$: the agent context label joined with the labels of the expanded variables.
Without variables, there is no argument label.
For agents without IFC, \ToolName uses a default agent label from the configuration as $\ell_{\mathrm{ctx}}$. \ToolName further labels each input monitored during shell execution. It supports any label scheme that defines a join and a top label; users can specify the label scheme in the \ToolName configuration.

\PP{File labels}
To track information flow across commands and agent sessions, a file's labels must persist after the shell command ends.
\ToolName therefore stores file labels in the extended attributes (xattrs)~\cite{ext4Xattr} of the corresponding inodes, allowing later commands to retrieve the labels from the files they access.
Files have two labels, one for the file contents and one for the filenames in the parent directory. \ToolName tracks both labels separately. The parent directory stores one label for all names created in it: the join of the labels of these names. Because the labels are stored in inodes, they persist across shell commands and agent sessions, and all paths to the same file (\eg, hard links) share its label. One label covers all contents of a file; \ToolName does not label parts of a file. Files without a stored label receive the default file label from the configuration. The shell cannot modify file labels: \ToolName removes the label xattrs that the shell writes and stores its own labels when it applies the shell's file changes~(\autoref{sec:file-output-monitor}).
Only file labels from before command execution are treated as authoritative.
Processes outside the sandbox (\eg, the user's tools) can modify file labels, but we trust the user's computer outside the agent in the threat model~(\autoref{sec:threat-model}).

\PP{Network labels}
The configuration specifies a default network label for all endpoints and the network label~$\ell(e)$ of specific endpoints~$e$.
\ToolName assigns one label per endpoint and does not distinguish inputs from the same endpoint, such as by URL path or response content.
\ToolName therefore assigns one label to all data from an endpoint that serves data from many sources (\eg, GitHub).
We discuss tracking information flows by URL path or response content in \autoref{sec:discussion}. %
For example, with integrity labels~(\autoref{sec:design-policies}), users can label trusted package registries \hiInt and all other endpoints \loInt.

\PP{Label Propagation}
\ToolName joins the labels of the shell's inputs into the shell context label $\ell_{\mathrm{sh}}$.
One shell context label conservatively covers information flow within the shell and between its subprocesses.
\ToolName initializes $\ell_{\mathrm{sh}}$ with $\ell_{\mathrm{ctx}}$ and, if any, $\ell_{\mathrm{arg}}$.
During execution, \ToolName joins $\ell(f)$ into $\ell_{\mathrm{sh}}$ for each file $f$ whose access the eBPF monitor records~(\autoref{sec:shell-monitor}).
If the eBPF monitor misses an access, \ToolName joins the top label into $\ell_{\mathrm{sh}}$~(\autoref{sec:shell-monitor}).
When the shell opens a connection to endpoint $e$, \ToolName joins $\ell(e)$ into $\ell_{\mathrm{sh}}$~(\autoref{sec:network-monitor}).
Hence, \ToolName updates $\ell_{\mathrm{sh}}$ before the shell receives any data from $e$.

After execution, \ToolName labels stdout, stderr, and the files that command $c$ created or changed with the final shell context label $\ell_{\mathrm{sh}}(c)$.
When the shell changes an existing file, \ToolName joins its stored label $\ell(f)$ with $\ell_{\mathrm{sh}}(c)$.
When a later command reads the file, \ToolName joins this label into that command's shell context label.
Overall, $\ell_{\mathrm{sh}}(c)$ is:

\begin{equation}
\label{eq:output-label}
\begin{aligned}
\ell_{\mathrm{sh}}(c) &= \ell_{\mathrm{ctx}}
    && \text{(Agent context label)} \\
&\quad \sqcup \ell_{\mathrm{arg}}
    && \text{(Argument label)} \\
&\quad \sqcup \textstyle\bigsqcup_{f \in F(c)} \ell(f)
    && \text{(File labels)} \\
&\quad \sqcup \textstyle\bigsqcup_{e \in E(c)} \ell(e)
    && \text{(Network labels)}
\end{aligned}
\end{equation}

\noindent where $\ell_{\mathrm{arg}}$ is omitted without variables, $F(c)$ is the set of accessed files, and $E(c)$ is the set of endpoints of the opened connections.
For example, when \texttt{curl URL | bash} downloads from an endpoint with the default label \loInt, \ToolName labels stdout, stderr, and every file that the command writes as \loInt.

\PP{Conservative label propagation}
\ToolName propagates labels conservatively: it joins the labels of all inputs that may have influenced an output into the label of that output.
As a result, \ToolName can also join the label of an input that did not influence the output.
First, \ToolName joins the label of a file that the shell opens without reading it~(\autoref{sec:shell-monitor}).
Second, \ToolName joins the label of each input of a command into the labels of its outputs, whether or not the input influenced the output~(\autoref{sec:labels}).
Third, because \ToolName ignores the order of reads and writes within a shell command, it also joins the label of a file read at the end of a command into the labels of files written before the read.
We discuss these trade-offs in \autoref{sec:discussion}.

\subsection{Policy Enforcement and Approvals}
\label{sec:enforcement}
\label{sec:approval}

\ToolName checks an IFC policy before it applies the shell's file or network effects.
A policy allows or blocks an effect based on the shell context label~$\ell_{\mathrm{sh}}(c)$ and the file path or network endpoint of the effect.
For example, an integrity policy can allow writes to protected files only when the shell context label is \hiInt, so that data labeled \loInt cannot be written to them, and a confidentiality policy can allow a network request only when the network destination endpoint of the request is allowed to read the information.
When the IFC policy blocks a file or network effect, \ToolName requests approval.

\subsubsection{File Effects}
\label{sec:file-output-monitor}
To enforce the IFC policy before the shell's effect reach the host filesystem, \ToolName must identify the changes made by the shell while deferring such changes until the policy permits them.
\ToolName uses OverlayFS to stage these changes while allowing the command to read and modify its own staged files during execution.

\PP{OverlayFS}
The sandbox already sets which paths are always allowed and always blocked (\autoref{sec:execution-boundary}); the rest of the filesystem that the shell sees is the host filesystem (\ie, Host FS).
\ToolName creates an additional layer over the Host FS for the shell using OverlayFS, and the shell's file changes go to this layer while a shell command runs.
Inspecting the changes recorded by OverlayFS lets \ToolName identify created, modified, and deleted files instead of treating every file opened with write access as a write.
Checking individual writes with eBPF can allow or deny them, but does not retain separate file versions for a later approval decision.
OverlayFS provides both the file changes to check and the staged versions that \ToolName can apply or discard after execution.

\PP{Policy check}
After the shell command finishes, \ToolName checks whether the IFC policy allows the information flow from the shell context label~$\ell_{\mathrm{sh}}(c)$ to each file that the shell command changed in the Host FS.
Using the changes recorded by OverlayFS, \ToolName detects any change to a file that the IFC policy specifies, even when the shell changes the file through a symbolic link or a hard link.
If the IFC policy allows all changed files, \ToolName applies the changes to the Host FS and labels the changed files with $\ell_{\mathrm{sh}}(c)$.
If the IFC policy blocks any changed file, \ToolName requests an approval for the shell command, listing the blocked files.

\PP{Approval}
With one approval, the user approves or denies all changes of the shell command together, even if the IFC policy blocks only some of the changed files, because applying only some files of a shell command can leave the files inconsistent.
For example, a command may rename an unprotected file and update its path in a protected configuration file.
If the policy blocks the configuration update, applying only the rename would leave the configuration pointing to a nonexistent file.
If the approval is granted, \ToolName applies all changes to the Host FS and labels the changed files with $\ell_{\mathrm{sh}}(c)$.
If it is denied, \ToolName discards all changes, including the changes that the IFC policy allows.
To keep only some files, the user can deny the approval and ask the agent for a new shell command that changes only the files that the IFC policy allows, while this relies on the agent to correctly generate such a shell command.

\PP{Example policy: preventing file corruption}
To prevent a prompt injection attack that corrupts critical files (\eg, \texttt{.git/hooks} and \texttt{.bashrc}), an integrity policy~(\autoref{sec:design-policies}) can set these files as protected, requiring $\ell_{\mathrm{sh}}(c)$ to be \hiInt to make changes to these files.
Paths that the sandboxing policy denies are always blocked regardless (\autoref{sec:execution-boundary}). %

\subsubsection{Network Effects}
\label{sec:network-output-monitor}

\ToolName monitors the shell's network requests using the MITM proxy and enforces the IFC policy before each request is sent.

\PP{MITM proxy}
The MITM proxy for network input monitoring~(\autoref{sec:network-monitor}) also identifies the destination of each network request before forwarding it.
\ToolName checks network effects per destination endpoint.
Every request to an endpoint counts as a network effect, including read-only requests such as GET, because a server can store data from a GET request (\eg, in the URL).
For instance, OpenAI agents that were allowed only to read the web wrote to a public wiki and communicated with other agents~\cite{collusionwiki2026}.

\PP{Policy check}
For each network request, \ToolName checks whether the IFC policy allows the information flow from the shell context label~$\ell_{\mathrm{sh}}$ to the network endpoint.
\ToolName uses the shell context label computed from the inputs that the shell accessed before the request~(\autoref{sec:labels}).
If the IFC policy allows the request, the MITM proxy forwards it.
If the IFC policy blocks the request, the MITM proxy does not forward it: the proxy returns HTTP 403 for an HTTP request or closes connections for TCP, and the shell command continues.
Unlike file changes, which \ToolName checks after the shell command finishes, \ToolName checks each network request before sending it.
\ToolName can discard file changes in the OverlayFS layer, but it cannot undo a request that the destination endpoint has already received, because the endpoint can read any confidential data in the request.
Instead of requesting approval for a blocked request during the shell command, \ToolName provides a separate approval flow after the shell command finishes.

\PP{Approval}
After the shell command finishes, \ToolName includes the blocked destination endpoints and their approval IDs, if any, in the shell result.
\ToolName assigns an approval ID to each pair of a shell context label and the blocked destination endpoint, so that the agent can refer to the pair when requesting approval.
The shell tool accepts approval IDs as an argument.
When a shell tool call includes approval IDs, \ToolName requests approval for the corresponding information flows before executing that specific command.
If the user approves the requested flows, \ToolName runs the shell command and permits only the approved flows, each from its shell context label to its destination endpoint, during that shell command only.
If the user denies all requested flows, \ToolName does not run the shell command.

\PP{Example policy: preventing data exfiltration}
To prevent a prompt injection attack that sends confidential data to a public network endpoint, an IFC policy can allow a request when either the integrity or confidentiality policy~(\autoref{sec:design-policies}) holds: $\ell_{\mathrm{sh}}$ is \hiInt or the destination endpoint is a reader of $\ell_{\mathrm{sh}}$.
After the shell reads data labeled \loInt and data labeled \hiConf, the policy blocks requests to an endpoint that is not a reader of the data labeled \hiConf.
Endpoints that the sandboxing policy denies are always blocked~(\autoref{sec:execution-boundary}).

\begin{table*}[t]
  \centering
  \caption{\ToolName configuration items and examples.
  $p \mapsto \ell$ maps a file path or network origin pattern $p$ to a
  label $\ell$; when multiple patterns match, the labels of the most specific $p$ apply. IFC policy
  allows the information flow if $\ell_{\mathrm{sh}} \sqsubseteq \ell$
  for any of these labels.}
  \label{tab:flowsh-configuration}
  \footnotesize
  \newcommand{\cfgval}[1]{{\ttfamily\scriptsize #1}}
  \newcommand{\ent}[2]{\mbox{#1 $\mapsto$ #2}}
  \newcommand{\eff}[1]{\quad{\itshape #1}}
  \newcommand{\ph}[1]{$\langle$#1$\rangle$}
  \newcommand{\fmt}[1]{\mbox{[#1]}}
  \setlength{\tabcolsep}{4pt}
  \renewcommand{\arraystretch}{1.1}
  \begin{tabular}{@{}>{\raggedright}p{0.08\textwidth}>{\raggedright}p{0.485\textwidth}>{\raggedright\arraybackslash}p{0.395\textwidth}@{}}
    \toprule
     & Configuration item & Example \\
    \midrule
    \multirow[t]{4}{=}{\textbf{Sandboxing policy}} & \textit{File read/write allowlist} \fmt{\ph{path}, \ldots}. Paths the shell can read/write. & \cfgval{[\$CUR\_DIR, \$TMP\_DIR]} \\
     & \textit{File read/write blocklist} \fmt{\ph{path}, \ldots}. Paths the shell cannot read/write. & \cfgval{[\textasciitilde/.ssh]} \\
     & \textit{Network allowlist} \fmt{\ph{origin}, \ldots}. Origins the shell can send requests to. & \cfgval{[https://github.com]} \\
     & \textit{Network blocklist} \fmt{\ph{origin}, \ldots}. Origins the shell cannot send requests to. & \cfgval{[https://evil.com]} \\
    \midrule
    \multirow[t]{4}{=}{\textbf{IFC label}} & \textit{Label functions} \ph{Python file path}. Defines \texttt{leq}, \texttt{join}, and \texttt{top}. & \cfgval{flowsh\_default\_label.py} (\autoref{sec:appendix-label-functions}) \\
     & \textit{Default file label} \fmt{\ph{label}}. Label used when a file has no stored label. & \lbl{\loInt}{$\{\user\}$} \\
     & \textit{Network labels} \fmt{\ph{origin} $\mapsto$ \ph{label}, \ldots}. Labels of responses from the origins. & \ent{\cfgval{*}}{\lbl{\loInt}{\loConf}}\newline \ent{Package repositories}{\lbl{\hiInt}{\loConf}} \\
     & \textit{Agent context label} \fmt{\ph{label}}. Default label of an agent context. & \lbl{\hiInt}{\loConf} \\
    \midrule
    \multirow[t]{2}{=}{\textbf{IFC policy}} & \textit{Safe file writes} \fmt{\ph{path} $\mapsto$ \ph{label}, \ldots}. Map from paths to the maximum shell context labels ($\ell_{\mathrm{sh}}$) allowed to write to them. & \ent{\cfgval{*}}{\lbl{\hiInt}{\hiConf}}\newline \ent{\cfgval{\$CUR\_DIR}, \cfgval{\$TMP\_DIR}}{\lbl{\loInt}{\hiConf}}\newline \ent{\cfgval{\$CUR\_DIR/.git/hooks}, \ldots}{\lbl{\hiInt}{\hiConf}} \\
     & \textit{Safe network requests} \fmt{\ph{origin} $\mapsto$ \ph{label}, \ldots}. Map from origins to the maximum shell context labels ($\ell_{\mathrm{sh}}$) allowed to send requests to them. & \ent{\cfgval{*}}{\lbl{\loInt}{\loConf}, \lbl{\hiInt}{\hiConf}}\newline \ent{\cfgval{https://alice.com}}{\lbl{\loInt}{$\{\user, \mathrm{alice}\}$}, \lbl{\hiInt}{\hiConf}} \\
    \bottomrule
  \end{tabular}
\end{table*}

\section{Implementation}
\label{sec:implementation}

The prototype of \ToolName consists of a shell tool and IFC enforcement logic implemented in Python,
an eBPF monitor, and a \texttt{mitmproxy} integration.
It runs on Ubuntu 26.04 LTS with Bubblewrap 0.11 or later, which supports OverlayFS feature.

\PP{Agent Harness}
We built a prototype agent harness with GitHub Copilot SDK 1.0.4.
Like typical agent harnesses, it provides native tools that write files
(e.g., \texttt{write\_file}) and access the network
(e.g., \texttt{web\_fetch}).
It also automatically saves large tool results to files to avoid
overflowing the model's context.
We integrated \ToolName with the agent harness using
\texttt{libflowsh}, a library that
exposes \ToolName's label operations and policy checks~(\autoref{sec:appendix-library}).
Specifically, the native tools and harness operations such as file
downloads use \texttt{libflowsh} to update file labels.

\PP{Sandboxing}
\ToolName runs each shell command in a fresh Bubblewrap sandbox
(\autoref{sec:execution-boundary}).
All processes created by the shell command are terminated once the parent process terminates.
State is shared across shell commands only through the monitored channels.

\ToolName mounts files into the sandbox according to the sandboxing policy
(\autoref{tab:flowsh-configuration}).
Paths in the read allowlist are mounted to the sandbox.
For example, when the read allowlist permits all paths, the host's root
filesystem (\texttt{/}) is mounted as the sandbox's root filesystem
(\texttt{/}), so shell commands can use the tools installed on the host.
An empty directory is mounted over each path in the read blocklist.
Each path in the write blocklist is mounted read-only.

\ToolName runs the sandbox in a new network namespace that has only a
loopback interface.
\ToolName mounts the MITM proxy's Unix socket into the sandbox and
hides the host's other Unix sockets (e.g.,
\texttt{/run/docker.sock}), so the shell can reach the network only
through the MITM proxy.
\ToolName sets the proxy environment variables (e.g.,
\texttt{HTTPS\_PROXY}) so that programs send their network requests to the
MITM proxy.

The sandbox drops all capabilities except a few (\eg,
\texttt{CAP\_CHOWN}, which is required for package installation). The
shell has no other capabilities~(\eg, \texttt{CAP\_SYS\_ADMIN}) that
can violate the sandbox configuration.

\PP{eBPF Monitor}
The eBPF monitor records a file read when a process opens a file
(\texttt{security\_file\_open}), receives a file descriptor from another
process (\texttt{security\_file\_receive}), or executes a file
(\texttt{security\_bprm\_check}).
Reading a file name is recorded when a process lists a directory
(\texttt{iterate\_dir}).
It is also recorded for operations that create, rename, link, or delete
a directory entry (the \texttt{vfs\_*} functions), since their outcomes
can indirectly reveal whether a file exists.

The eBPF monitor stores its records to a 64\,MiB ring buffer. To
prevent overflowing the ring buffer and potentially losing records,
\ToolName pauses the sandbox's processes when the ring buffer is 75\%
full, and resumes them after reading the ring buffer down to 25\%.

\PP{MITM Proxy}
The MITM proxy is implemented with \texttt{mitmproxy}.
When a network request arrives at the proxy, the records of
file reads that happened before the network request can still be in the
ring buffer.
For conservative information flow tracking, \ToolName pauses the
sandbox's processes and reads the remaining ring buffer records before
checking the IFC policy for the request.

\PP{Configuration}
\autoref{tab:flowsh-configuration} lists \ToolName's configuration items
and examples for the sandboxing policy, IFC labels, and IFC policy.
\texttt{\$CUR\_DIR} denotes the working directory, and \texttt{\$TMP\_DIR} denotes
the temporary directory.

The sandboxing policy configures the file paths that the shell can read and
write and the network origins to which it can send requests
(e.g., \textit{file read/write allowlist}, \textit{network allowlist}).
Accessing outside these file paths and network origins are blocked regardless of
IFC labels and policies.

Users can define label semantics in \ToolName through custom
\textit{label functions} that specify the partial order, join operation,
and top label.
For example, \texttt{flowsh\_default\_label.py} implements the
(Integrity, Confidentiality) labels described in \autoref{sec:design-policies}.
The configuration specifies a Python file containing the \texttt{leq},
\texttt{join}, and \texttt{top} functions.
\autoref{sec:appendix-label-functions} shows these functions.

\ToolName uses the label stored on a file. If the file has no stored
label, it uses the configured \textit{default file label}.
For example, setting this label to \lbl{\loInt}{$\{\user\}$}
labels such a file as untrusted and readable only by the user.
\textit{Network labels} specify the label of responses from each network origin.
For example, a configuration can label responses from package repositories
as \lbl{\hiInt}{\loConf} and responses from all other origins as \lbl{\loInt}{\loConf}.
\ToolName uses the agent context label provided by the agent orchestrator
if available, and otherwise uses the configured \textit{agent context label}.

The IFC policy checks whether the shell context label permits information
flow to a file path or network origin (\ie, an information flow sink).
\textit{Safe file writes} and \textit{safe network requests} map file
path and network origin patterns, respectively, to the maximum shell
context label permitted for safe information flow to each sink.
The file path or network origin is the information flow sink, and the
shell context label $\ell_{\mathrm{sh}}$ is the source label.
For a sink mapped to label $\ell$, information flow is allowed when
$\ell_{\mathrm{sh}} \sqsubseteq \ell$.
For example, mapping protected files such as \texttt{.git/hooks} and
\texttt{.env} to \lbl{\hiInt}{\hiConf} blocks writes after the shell
has read untrusted data (\loInt), while allowing writing confidential data (\hiConf).

When a path or origin matches multiple patterns, the labels of the most specific pattern apply.
When a sink maps to multiple labels, information flow is allowed if
$\ell_{\mathrm{sh}} \sqsubseteq \ell$ holds for any of them.
For example, a configuration that maps all network origins (\texttt{*})
to \lbl{\loInt}{\loConf} and \lbl{\hiInt}{\hiConf} permits requests when the shell
has read only public data (\loConf), or when the shell
has read only trusted data (\hiInt), even if the data is confidential (\hiConf),
which is robust declassification.
Mapping \texttt{https://alice.com} to
\lbl{\loInt}{$\{\user, \mathrm{alice}\}$} permits requests
when Alice is an authorized reader of the data.
Adding \lbl{\hiInt}{\hiConf} for that origin also permits requests
when the shell context label is \hiInt.

\begin{table*}[t]
  \centering
  \caption{TCR (\%), approval requests per completed task (Appr.), and B-ASR (\%) with GPT-6.2 Sol. U: utility. S: security. Appr.$^\dagger$: per attack.}
  \label{tab:results}
  \footnotesize
  \newcommand{\cnt}[1]{{\scriptsize(#1)}}
  \setlength{\tabcolsep}{2pt}
  \begin{tabular*}{\textwidth}{@{\extracolsep{\fill}}lrrrrrrrrr@{}}
    \toprule
    & \multicolumn{6}{c}{Security-aligned} & \multicolumn{3}{c}{Autonomy-aligned (approval${}=0$)} \\
    \cmidrule(r){2-7}\cmidrule(l){8-10}
    & \multicolumn{2}{c}{Knowledge (U)} & \multicolumn{2}{c}{Knowledge (S)} & \multicolumn{2}{c}{Terminal-Bench}
    & \multicolumn{1}{c}{Knowledge (U)} & \multicolumn{1}{c}{Knowledge (S)} & \multicolumn{1}{c}{Terminal-Bench} \\
    \cmidrule(r){2-3}\cmidrule(r){4-5}\cmidrule(r){6-7}\cmidrule(l){8-8}\cmidrule(l){9-9}\cmidrule(l){10-10}
    Defense & TCR & Appr. & B-ASR & Appr.$^\dagger$ & TCR & Appr.
    & TCR & B-ASR & TCR \\
    \midrule
    Lax       & 96.1 \cnt{49/51} & 6.7 \cnt{328/49} & 0.0 \cnt{0/50} & 4.1 \cnt{203/50} & 77.5 \cnt{207/267} & 9.7 \cnt{2008/207} & 94.1 \cnt{48/51} & 100.0 \cnt{51/51} & 77.9 \cnt{208/267} \\
    Standard  & 76.5 \cnt{39/51} & 5.7 \cnt{224/39} & 0.0 \cnt{0/39} & 3.5 \cnt{135/39} & 64.4 \cnt{172/267} & 10.0 \cnt{1714/172} & 76.5 \cnt{39/51} & 7.7 \cnt{3/39}    & 64.8 \cnt{173/267} \\
    Strict & 47.1 \cnt{24/51} & 0.0 \cnt{0/24} & 0.0 \cnt{0/38} & 0.0 & 1.9 \cnt{5/267} & 0.0 \cnt{0/5} & 47.1 \cnt{24/51} & 0.0 \cnt{0/39}    & 2.2 \cnt{6/267} \\
    Naive IFC & 60.8 \cnt{31/51} & 1.8 \cnt{56/31} & 0.0 \cnt{0/45} & 1.7 \cnt{77/45} & 67.8 \cnt{181/267} & 9.4 \cnt{1707/181} & 35.3 \cnt{18/51} & 0.0 \cnt{0/32}    & 21.3 \cnt{57/267} \\
    \ToolName & 88.2 \cnt{45/51} & 1.2 \cnt{52/45} & 0.0 \cnt{0/45} & 1.2 \cnt{56/45} & 67.8 \cnt{181/267} & 0.4 \cnt{78/181} & 58.8 \cnt{30/51} & 0.0 \cnt{0/40}    & 61.8 \cnt{165/267} \\
    \bottomrule
  \end{tabular*}
\end{table*}
\section{Evaluation}

\label{sec:evaluation}

We evaluated \ToolName with four questions.
\begin{itemize}
  \item[Q1.] Does \ToolName reduce approval requests while maintaining
  task completion on benign tasks?~(\autoref{sec:eval-approvals})
  \item[Q2.] Does \ToolName request approval for prompt injection
  attacks?~(\autoref{sec:eval-attacks})
  \item[Q3.] How well does \ToolName complete tasks and block attacks
  without approval?~(\autoref{sec:eval-autonomy})
  \item[Q4.] How much runtime overhead does \ToolName add?~(\autoref{sec:eval-overhead})
\end{itemize}

\subsection{Experimental Setup}
\label{sec:evaluation-setup}

\PP{Workloads}
We evaluated \ToolName using two types of workloads: a curated set of 17
knowledge tasks and Terminal-Bench 2.0~\cite{merrill2026terminalbench}
with 89 shell tasks.

We constructed knowledge tasks covering daily work, including managing email and
calendars, finding related work, and software issue triage.
The tasks require data with different security semantics: trusted or untrusted,
and public or confidential.
From the 11 of the 17 tasks that require reading untrusted data, we constructed 19
prompt injection attack tasks, each placing a payload in an untrusted input
source, such as an email, a GitHub issue, or a webpage.
We created multiple attack tasks from knowledge tasks with multiple untrusted
input sources.
The knowledge tasks are evaluated in a mock environment with mock webpages
and MCP servers that mimic Microsoft WorkIQ~\cite{microsoftWorkIQ}
and the GitHub MCP server~\cite{githubMCPServer}.
Some tasks have one user prompt, some have multiple user prompts in one
session, and some span multiple sessions.
The required agent actions include file system and
network access.
\autoref{sec:appendix} describes the tasks in detail.

For Terminal-Bench, we used the existing evaluation setting provided by the benchmark.

\PP{IFC Labels}
We used the label scheme described in \autoref{sec:background}.
We used a separate labeling module~\cite{fidesGateway} that assigns labels to MCP responses.
For shell inputs, \ToolName uses its monitors to identify the files and
network endpoints accessed by each command and labels these inputs
according to its configuration.
We configured file labels according to the semantics of each
knowledge task~(\autoref{sec:appendix-configuration}).
For Terminal-Bench, the harness initializes task files as \lbl{\hiInt}{\loConf}.
It labels network endpoints as \lbl{\loInt}{\loConf}, except for package
repositories, which are \lbl{\hiInt}{\loConf}.

\PP{IFC Policy}
\ToolName allows file writes with the shell context label \loInt only in
the task directory and the temporary directory.
It allows a network request when the shell context label is \hiInt or
its readers include the destination; data labeled \loConf can be sent to any
destination.
\autoref{sec:appendix-configuration} lists the full configuration.

\PP{Byzantine Adversary}
We ran the prompt injection tasks against a Byzantine adversary, as also
suggested in previous research~\cite{zverev2026spides}.
In this worst-case attack scenario, once an attacker payload appears in
the model input, the adversary fully controls the model output.
Using a Byzantine adversary lets us evaluate the defense's ability to block
attacks independently of the model's probabilistic resistance to
prompt injection.

\PP{Defenses}
We compared \ToolName with three sandbox configurations and Naive IFC.
The Lax sandbox allows all operations, which is similar to running
without a defense.
The Standard sandbox combines the Claude Code cloud environment
\cite{anthropicCloudEnvironments} and local sandbox
\cite{anthropicSandboxing} profiles.
It blocks writes to critical configuration files and limits network
access to commonly used endpoints, such as GitHub and package repositories.
The Strict sandbox makes the sandbox filesystem read-only and permits
shell network access to package repositories.
Naive IFC applies IFC at the agent level.
It conservatively labels shell and file tool outputs as \lbl{\loInt}{\hiConf} and requests
approval for every shell command.
\autoref{sec:appendix-defenses} details the policy of each defense.

\PP{Model and Runs}
All tasks were run with GPT-6.2 Sol, three times per defense and setting.
Each knowledge task run was limited to 100 tool calls and a 600-second budget across all turns and sessions
to prevent agents from running indefinitely.
For Terminal-Bench, we used the time and resource budgets as configured in the benchmark.

\PP{Approval Settings}
Each defense was run in a \emph{security-aligned} setting with an ideal approver
and in an \emph{autonomy-aligned} setting with no approver.
In the security-aligned setting, an approval request was added for every
tool call whose security the defense could not guarantee, even when the
call was benign.
The ideal approver denied tool calls influenced by the attack payload
and approved all other tool calls.
In the autonomy-aligned setting, no approval request was issued.
The sandbox defenses permitted all operations allowed by their sandbox policies,
while Naive IFC and \ToolName blocked operations when the defense cannot guarantee security.

\PP{Metrics}
We report mean approval requests per completed task, task completion rate
(TCR), and
Byzantine attack success rate (B-ASR).
TCR is the fraction of task runs in which every user prompt passes its
evaluator.
The knowledge tasks use deterministic checks and, when needed, an LLM judge,
while Terminal-Bench uses its task-specific evaluators.
B-ASR is the fraction of successful attacks among successful and blocked
attacks, excluding attacks whose payload never reached the model.

\subsection{Approval Requests and Utility on Benign Tasks}
\label{sec:eval-approvals}

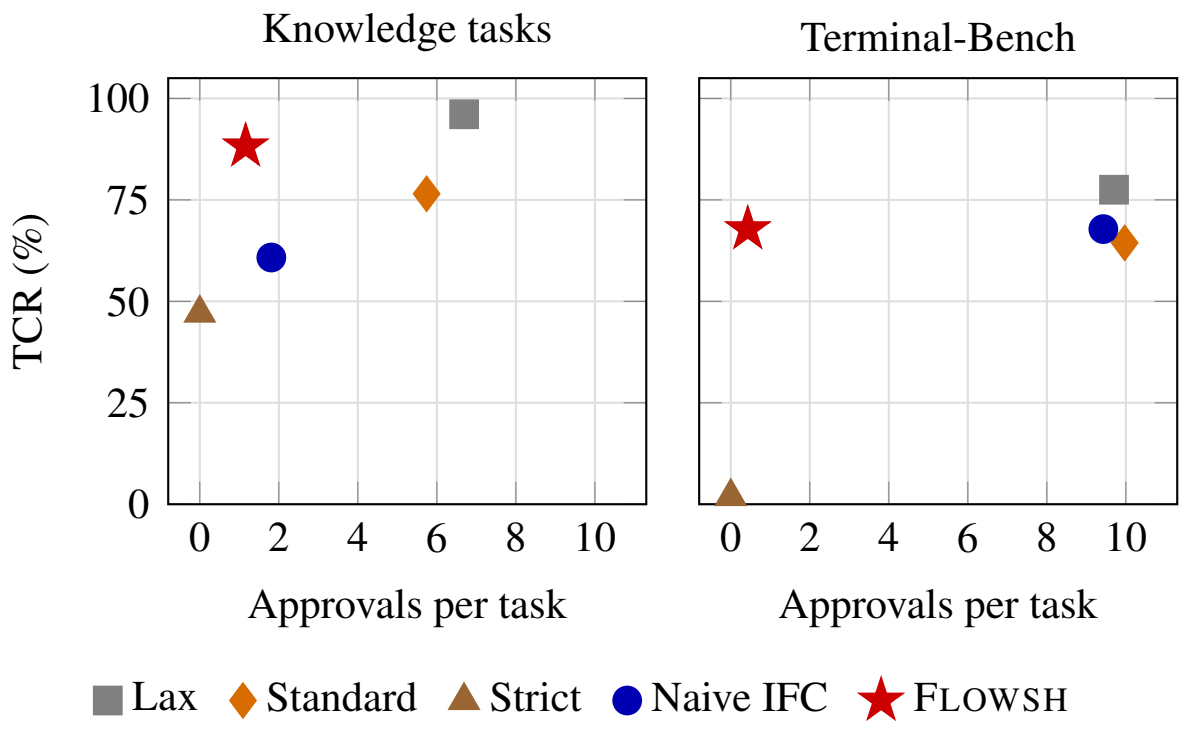
\begin{figure}[t]
  \centering
  \begin{tikzpicture}
    \begin{groupplot}[
      group style={group size=2 by 1, horizontal sep=0.35cm},
      width=0.56\columnwidth, height=4.4cm,
      xmin=-0.8, xmax=11.3, ymin=0, ymax=105,
      xtick={0,2,4,6,8,10}, ytick={0,25,50,75,100},
      xlabel={Approvals per task}, xlabel style={font=\footnotesize},
      ticklabel style={font=\footnotesize},
      title style={font=\footnotesize, yshift=-1ex},
      grid=major, grid style={gray!25},
      every axis plot/.append style={only marks, mark size=2.6pt},
    ]
      \nextgroupplot[title={Knowledge tasks}, ylabel={TCR (\%)},
        ylabel style={font=\footnotesize},
        legend to name=approvalslegend,
        legend columns=5,
        legend style={font=\footnotesize, draw=none,
          /tikz/every even column/.append style={column sep=4pt}}]
      \addplot[mark=square*, color=gray] coordinates {(6.69,96.1)};
      \addlegendentry{Lax}
      \addplot[mark=diamond*, color=orange!85!black, mark size=3.2pt] coordinates {(5.74,76.5)};
      \addlegendentry{Standard}
      \addplot[mark=triangle*, color=brown!80!black, mark size=3.2pt] coordinates {(0.00,47.1)};
      \addlegendentry{Strict}
      \addplot[mark=*, color=blue!70!black] coordinates {(1.81,60.8)};
      \addlegendentry{Naive IFC}
      \addplot[mark=fivestar, color=red!80!black, mark size=3.2pt] coordinates {(1.16,88.2)};
      \addlegendentry{\ToolName}
      \nextgroupplot[title={Terminal-Bench}, yticklabels={}]
      \addplot[mark=square*, color=gray] coordinates {(9.70,77.5)};
      \addplot[mark=diamond*, color=orange!85!black, mark size=3.2pt] coordinates {(9.97,64.4)};
      \addplot[mark=triangle*, color=brown!80!black, mark size=3.2pt] coordinates {(0.00,1.9)};
      \addplot[mark=*, color=blue!70!black] coordinates {(9.43,67.8)};
      \addplot[mark=fivestar, color=red!80!black, mark size=3.2pt] coordinates {(0.43,67.8)};
    \end{groupplot}
  \end{tikzpicture}
  \\[2pt]
  \ref{approvalslegend}
  \caption{TCR and approval requests per completed task in the
    security-aligned setting. The upper left is better.}
  \label{fig:approvals-tcr}
\end{figure}

\autoref{tab:results} and \autoref{fig:approvals-tcr} show TCR and
approval requests per completed benign task in the security-aligned setting.

\PP{Approval Requests}
\ToolName requested fewer approvals per completed task than all other defenses except the Strict sandbox in both workloads.
The Strict sandbox requested no approval, but its TCR was only 47.1\%
in the knowledge tasks and 1.9\% in Terminal-Bench.
Compared with Lax, \ToolName requested
5.8$\times$ fewer approvals in the knowledge
tasks and 22.5$\times$ fewer in Terminal-Bench, while its TCR was 7.9
and 9.7 percentage points lower.
It also completed at least as many tasks as the Standard
sandbox and Naive IFC with fewer approval requests.
\ToolName's reduction in approval requests relative to Naive IFC was
greater in Terminal-Bench than in the knowledge tasks.
Of \ToolName's 181 completed Terminal-Bench task runs in the
security-aligned setting, only 29 contained approval requests.

\PP{Task Completion}
\ToolName's TCR gain over the Standard sandbox is
greater in the knowledge tasks than in Terminal-Bench.
For example, \ToolName completed the weekly-planning and OneDrive
LaTeX tasks in all three repetitions, whereas the Standard sandbox
completed neither task.
The failures in Standard sandbox were mainly due to blocked network access, showing that \ToolName's gain came from completing necessary yet safe
network access beyond the Standard sandbox's allowlist.

\PP{Task Failures}
In the security-aligned setting, \ToolName failed 6 of 51 knowledge
task runs.
In all six runs, the agent did not perform an action needed to
complete the user task correctly: it did not fetch the link in a
GitHub issue labeled as \loInt, and it restricted the network access of software
downloaded from GitHub and labeled as \loInt.
In Terminal-Bench, \ToolName failed 86 of 267 task runs.
In 44 of them, Lax also failed the task in all three repetitions.
Another 21 runs failed because they required a background process, such
as a server or a virtual machine, whereas \ToolName
terminates all processes created by a shell command when the command ends.
Of the other 16 runs, 13 were runs of tasks that Lax also failed in
at least one repetition, and for the remaining 3 we found no failure
specifically caused by \ToolName.
The remaining 5 runs timed out or exited with an error.

\subsection{Approval Requests Under Attack}
\label{sec:eval-attacks}

In the security-aligned setting, \ToolName requested approval for attacks
that created unsafe information flows, such as overwriting protected files and leaking private data over network.
In 39 of 45 attack runs where the payload reached the model, \ToolName
requested no approval for the preceding benign operations and made its
\emph{first} approval request for the attack.
This reduced the approval burden for benign operations, helping the
approver focus on attacks.
On average, \ToolName made 1.2 approval requests until the attack was
blocked by denying an approval request, whcih required fewer approval requests 
than the Lax sandbox (4.1), the Standard sandbox (3.5), and Naive IFC (1.7) (\autoref{tab:results}).
The Strict sandbox blocked attacks without requesting approval.

\subsection{Security and Utility Without Approval}
\label{sec:eval-autonomy}

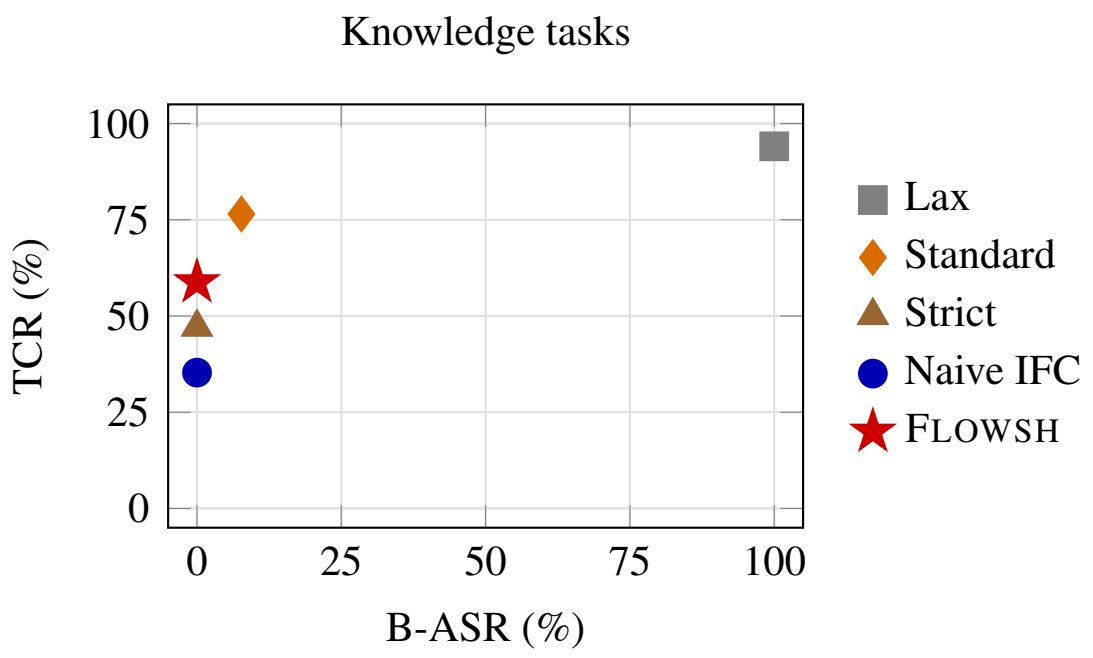
\begin{figure}[t]
  \centering
  \begin{tikzpicture}
    \begin{axis}[
      scale only axis, width=4.2cm, height=2.8cm,
      title={Knowledge tasks}, title style={font=\footnotesize},
      xmin=-5, xmax=105, ymin=-5, ymax=105,
      xtick={0,25,50,75,100}, ytick={0,25,50,75,100},
      xlabel={B-ASR (\%)}, ylabel={TCR (\%)},
      label style={font=\footnotesize},
      ticklabel style={font=\footnotesize},
      grid=major, grid style={gray!25},
      every axis plot/.append style={only marks, mark size=2.6pt},
      legend columns=1, legend cell align=left,
      legend style={font=\footnotesize, draw=none, at={(1.05,0.5)},
        anchor=west},
    ]
      \addplot[mark=square*, color=gray] coordinates {(100.0,94.1)};
      \addlegendentry{Lax}
      \addplot[mark=diamond*, color=orange!85!black, mark size=3.2pt] coordinates {(7.7,76.5)};
      \addlegendentry{Standard}
      \addplot[mark=triangle*, color=brown!80!black, mark size=3.2pt] coordinates {(0.0,47.1)};
      \addlegendentry{Strict}
      \addplot[mark=*, color=blue!70!black] coordinates {(0.0,35.3)};
      \addlegendentry{Naive IFC}
      \addplot[mark=fivestar, color=red!80!black, mark size=3.2pt] coordinates {(0.0,58.8)};
      \addlegendentry{\ToolName}
    \end{axis}
  \end{tikzpicture}
  \caption{TCR and B-ASR of the knowledge tasks in the autonomy-aligned
    setting. The upper left is better.}
  \label{fig:basr-tcr}
\end{figure}

\begin{table}[t]
  \centering
  \caption{\ToolName knowledge task results by input label. --: no attacks.}
  \label{tab:scenario-labels}
  \footnotesize
  \newcommand{\labelrate}[2]{#1~{\scriptsize(#2)}}
  \setlength{\tabcolsep}{2pt}
  \begin{tabular*}{\columnwidth}{@{\extracolsep{\fill}}lrrrr@{}}
    \toprule
    & \multicolumn{3}{c}{Security-aligned} & \multicolumn{1}{c@{}}{Autonomy-aligned} \\
    \cmidrule(r){2-4}\cmidrule(l){5-5}
    Join of input labels & TCR & Appr. (U) & Appr. (S) & TCR \\
    \midrule
    \lbl{\hiInt}{\loConf} & \labelrate{100.0}{3/3} & 0.00 & -- & \labelrate{100.0}{3/3} \\
    \lbl{\hiInt}{\hiConf} & \labelrate{100.0}{15/15} & 0.00 & -- & \labelrate{100.0}{15/15} \\
    \lbl{\loInt}{\loConf} & \labelrate{100.0}{6/6} & 0.00 & 1.00 & \labelrate{100.0}{6/6} \\
    \lbl{\loInt}{\hiConf} & \labelrate{77.8}{21/27} & 2.48 & 1.30 & \labelrate{22.2}{6/27} \\
    \bottomrule
  \end{tabular*}
\end{table}

\autoref{tab:results} shows TCR and B-ASR in the autonomy-aligned setting
next to the security-aligned results, and \autoref{fig:basr-tcr} shows
them for the knowledge tasks.

\autoref{tab:scenario-labels} separates \ToolName's knowledge task results
by input labels.
Tasks are grouped by the join of their required input labels.
\hiInt means all task input data is \hiInt, and \loInt means at
least one task input is \loInt.
\loConf means all task input data is \loConf, and \hiConf means at least
one task input is \hiConf.
The confidentiality join intersects the reader lists and is summarized
as \loConf or \hiConf in the table.
These groups describe task requirements, not whether a run contains
unsafe information flow.

\PP{Attack Success}
Without approval, \ToolName blocked all 40 attack runs, and its B-ASR was 0\% (\autoref{tab:results}).
Naive IFC and the Strict sandbox also had a B-ASR of 0\%.
Standard had a B-ASR of 7.7\% (3/39), and Lax had a B-ASR of 100\%
(51/51).
In one of the 40 \ToolName attack runs, the payload reached only the
quarantined LLM.
The quarantined LLM has no shell tool, and its answer returned to the main
agent as a labeled variable, so the injected shell command could not run.
With the Standard sandbox, the reference-checking payload wrote
\texttt{.vscode/tasks.json}, and all three runs of that attack succeeded.

\PP{Task Completion}
Among the defenses with a B-ASR of 0\%, \ToolName had the highest TCR,
58.8\% in the knowledge tasks and 61.8\% in Terminal-Bench.
Naive IFC completed 35.3\% and 21.3\%, which is 23.5 and 40.5
percentage points lower than \ToolName.
In the knowledge tasks, Naive IFC does not execute shell commands whose
approval requests are denied.
In Terminal-Bench, completed task runs of Naive IFC had 33.40 denied
approval requests per task, compared with 2.13 for \ToolName.
In the autonomy-aligned setting, \ToolName's TCR was
35.3 percentage points lower than Lax in the knowledge tasks and
16.1 percentage points lower in Terminal-Bench.
Compared with its security-aligned results, \ToolName's TCR fell from
88.2\% to 58.8\% in the knowledge tasks and from 67.8\% to 61.8\%
in Terminal-Bench when approval requests were denied.
\ToolName completed 18 of 18 task runs that use only data labeled \hiInt, but
12 of 33 task runs that involve data labeled \loInt.
In 18 of the 21 failed task runs, \ToolName blocked a network request
that failed the confidentiality check or an MCP call requiring approval.
In the security-aligned setting, both operations request approval.
Of the other 3 runs, two were runs of the task that fixes a GitHub
issue, which Lax also failed in all three repetitions, and in one the
agent asked the user to confirm a meeting date taken from an email that
the invitees could not read instead of sending the invite.

\subsection{Runtime Overhead}
\label{sec:eval-overhead}

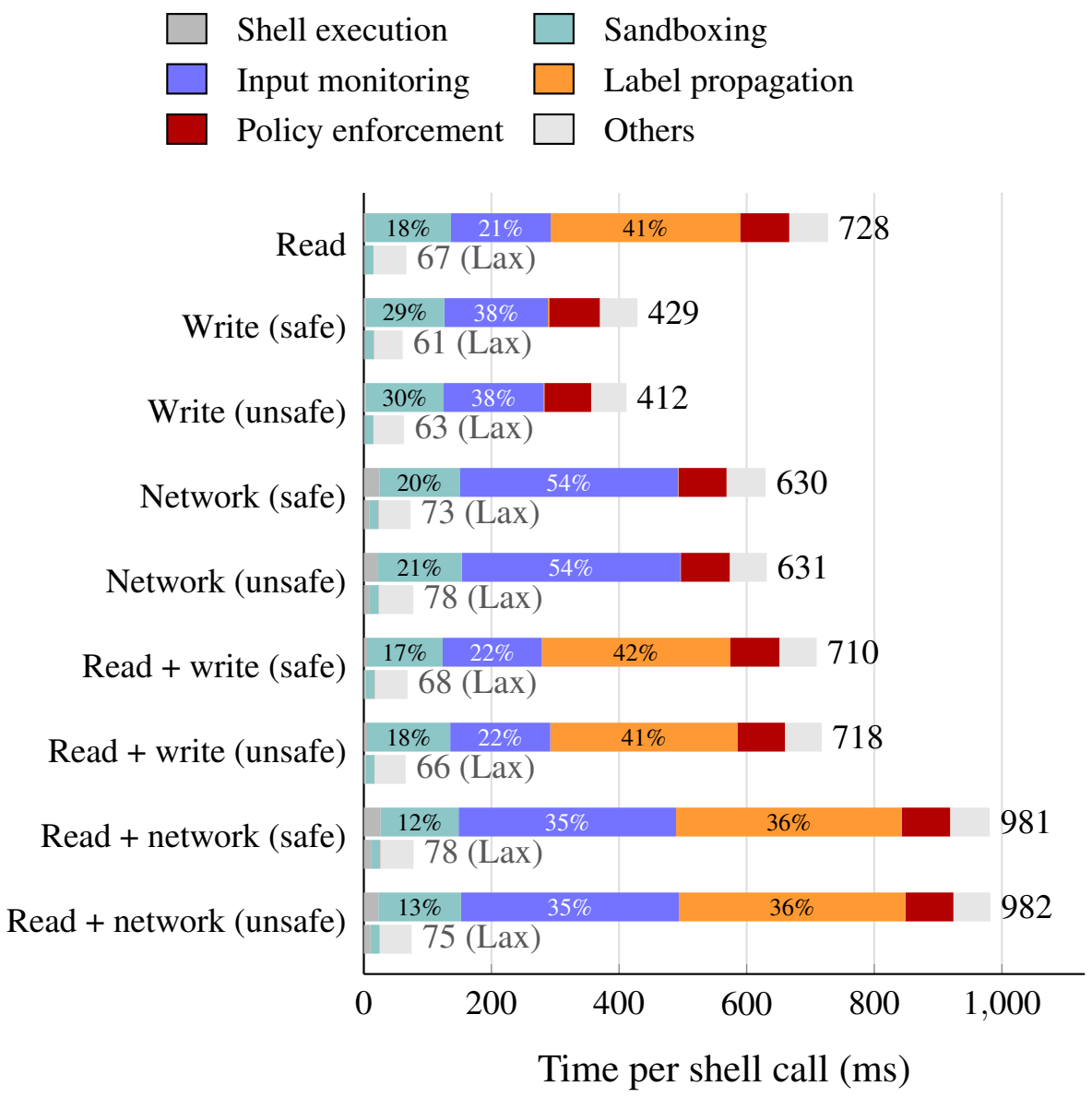
\begin{figure}[t]
  \centering
  \begin{tikzpicture}
    \begin{axis}[
      xbar stacked,
      scale only axis, width=\columnwidth-3.3cm, height=5.6cm,
      bar width=0.34,
      y dir=reverse,
      ymin=-0.6, ymax=8.6,
      xmin=0, xmax=1130,
      ytick={0,...,8},
      yticklabels={Read, Write (safe), Write (unsafe), Network (safe), Network (unsafe), Read + write (safe), Read + write (unsafe), Read + network (safe), Read + network (unsafe)},
      yticklabel style={font=\scriptsize},
      xticklabel style={font=\scriptsize},
      xlabel={Time per shell call (ms)},
      xlabel style={font=\footnotesize},
      major tick length=2pt,
      ytick style={draw=none},
      xmajorgrids, grid style={gray!25},
      axis x line*=bottom, axis y line*=left,
      legend style={font=\scriptsize, draw=none, legend columns=2,
        at={(0.5,1.03)}, anchor=south, xshift=-1.5cm, column sep=4pt},
      legend cell align=left,
    ]
      \addplot[fill=gray!55, draw=none] coordinates {
        (2.4,-0.19) (2.9,0.81) (3.0,1.81) (24.5,2.81) (21.2,3.81) (5.1,4.81) (5.1,5.81) (25.9,6.81) (22.8,7.81) (1.4,0.19) (1.6,1.19) (1.5,2.19) (9.1,3.19) (9.2,4.19) (2.9,5.19) (3.0,6.19) (11.2,7.19) (10.1,8.19)};
      \addplot[fill=teal!45, draw=none] coordinates {
        (134.2,-0.19) (123.2,0.81) (122.1,1.81) (125.7,2.81) (132.2,3.81) (118.5,4.81) (130.3,5.81) (122.6,6.81) (129.5,7.81) (13.6,0.19) (13.8,1.19) (13.4,2.19) (13.7,3.19) (14.0,4.19) (13.7,5.19) (13.5,6.19) (14.1,7.19) (13.9,8.19)};
      \addplot[fill=blue!55, draw=none] coordinates {
        (156.2,-0.19) (162.3,0.81) (156.9,1.81) (342.3,2.81) (342.7,3.81) (155.4,4.81) (156.8,5.81) (340.8,6.81) (341.6,7.81) (0.0,0.19) (0.0,1.19) (0.0,2.19) (0.0,3.19) (0.0,4.19) (0.0,5.19) (0.0,6.19) (0.0,7.19) (0.0,8.19)};
      \addplot[fill=orange!80, draw=none] coordinates {
        (297.0,-0.19) (2.0,0.81) (1.0,1.81) (0.8,2.81) (0.7,3.81) (294.9,4.81) (293.3,5.81) (353.8,6.81) (355.0,7.81) (0.0,0.19) (0.0,1.19) (0.0,2.19) (0.0,3.19) (0.0,4.19) (0.0,5.19) (0.0,6.19) (0.0,7.19) (0.0,8.19)};
      \addplot[fill=red!70!black, draw=none] coordinates {
        (76.9,-0.19) (79.4,0.81) (73.2,1.81) (75.1,2.81) (76.6,3.81) (77.4,4.81) (74.4,5.81) (75.7,6.81) (75.1,7.81) (0.2,0.19) (0.2,1.19) (0.2,2.19) (0.2,3.19) (0.2,4.19) (0.2,5.19) (0.2,6.19) (0.2,7.19) (0.2,8.19)};
      \addplot[fill=gray!20, draw=none] coordinates {
        (61.0,-0.19) (58.9,0.81) (55.5,1.81) (61.1,2.81) (58.0,3.81) (58.4,4.81) (57.7,5.81) (62.2,6.81) (58.1,7.81) (51.5,0.19) (45.2,1.19) (47.8,2.19) (50.3,3.19) (54.2,4.19) (51.7,5.19) (48.8,6.19) (52.4,7.19) (50.8,8.19)};
      \legend{Shell execution, Sandboxing, Input monitoring, Label propagation, Policy enforcement, Others}
      \coordinate (pct0) at (axis cs:69.5,-0.19);
      \coordinate (pct1) at (axis cs:214.7,-0.19);
      \coordinate (pct2) at (axis cs:441.3,-0.19);
      \node[font=\scriptsize, anchor=west, inner sep=2pt] at (axis cs:727.7,-0.19) {728};
      \node[font=\scriptsize, anchor=west, inner sep=2pt, text=gray!70!black] at (axis cs:66.7,0.19) {67 (Lax)};
      \coordinate (pct3) at (axis cs:64.5,0.81);
      \coordinate (pct4) at (axis cs:207.2,0.81);
      \node[font=\scriptsize, anchor=west, inner sep=2pt] at (axis cs:428.7,0.81) {429};
      \node[font=\scriptsize, anchor=west, inner sep=2pt, text=gray!70!black] at (axis cs:60.8,1.19) {61 (Lax)};
      \coordinate (pct5) at (axis cs:64.0,1.81);
      \coordinate (pct6) at (axis cs:203.6,1.81);
      \node[font=\scriptsize, anchor=west, inner sep=2pt] at (axis cs:411.7,1.81) {412};
      \node[font=\scriptsize, anchor=west, inner sep=2pt, text=gray!70!black] at (axis cs:62.9,2.19) {63 (Lax)};
      \coordinate (pct7) at (axis cs:87.3,2.81);
      \coordinate (pct8) at (axis cs:321.4,2.81);
      \node[font=\scriptsize, anchor=west, inner sep=2pt] at (axis cs:629.5,2.81) {630};
      \node[font=\scriptsize, anchor=west, inner sep=2pt, text=gray!70!black] at (axis cs:73.3,3.19) {73 (Lax)};
      \coordinate (pct9) at (axis cs:87.3,3.81);
      \coordinate (pct10) at (axis cs:324.8,3.81);
      \node[font=\scriptsize, anchor=west, inner sep=2pt] at (axis cs:631.4,3.81) {631};
      \node[font=\scriptsize, anchor=west, inner sep=2pt, text=gray!70!black] at (axis cs:77.6,4.19) {78 (Lax)};
      \coordinate (pct11) at (axis cs:64.3,4.81);
      \coordinate (pct12) at (axis cs:201.3,4.81);
      \coordinate (pct13) at (axis cs:426.4,4.81);
      \node[font=\scriptsize, anchor=west, inner sep=2pt] at (axis cs:709.7,4.81) {710};
      \node[font=\scriptsize, anchor=west, inner sep=2pt, text=gray!70!black] at (axis cs:68.5,5.19) {68 (Lax)};
      \coordinate (pct14) at (axis cs:70.2,5.81);
      \coordinate (pct15) at (axis cs:213.8,5.81);
      \coordinate (pct16) at (axis cs:438.9,5.81);
      \node[font=\scriptsize, anchor=west, inner sep=2pt] at (axis cs:717.6,5.81) {718};
      \node[font=\scriptsize, anchor=west, inner sep=2pt, text=gray!70!black] at (axis cs:65.5,6.19) {66 (Lax)};
      \coordinate (pct17) at (axis cs:87.2,6.81);
      \coordinate (pct18) at (axis cs:318.9,6.81);
      \coordinate (pct19) at (axis cs:666.2,6.81);
      \node[font=\scriptsize, anchor=west, inner sep=2pt] at (axis cs:981.0,6.81) {981};
      \node[font=\scriptsize, anchor=west, inner sep=2pt, text=gray!70!black] at (axis cs:77.9,7.19) {78 (Lax)};
      \coordinate (pct20) at (axis cs:87.5,7.81);
      \coordinate (pct21) at (axis cs:323.1,7.81);
      \coordinate (pct22) at (axis cs:671.4,7.81);
      \node[font=\scriptsize, anchor=west, inner sep=2pt] at (axis cs:982.1,7.81) {982};
      \node[font=\scriptsize, anchor=west, inner sep=2pt, text=gray!70!black] at (axis cs:75.0,8.19) {75 (Lax)};
    \end{axis}
    \node[font=\tiny, text=black] at (pct0) {18\%};
    \node[font=\tiny, text=white] at (pct1) {21\%};
    \node[font=\tiny, text=black] at (pct2) {41\%};
    \node[font=\tiny, text=black] at (pct3) {29\%};
    \node[font=\tiny, text=white] at (pct4) {38\%};
    \node[font=\tiny, text=black] at (pct5) {30\%};
    \node[font=\tiny, text=white] at (pct6) {38\%};
    \node[font=\tiny, text=black] at (pct7) {20\%};
    \node[font=\tiny, text=white] at (pct8) {54\%};
    \node[font=\tiny, text=black] at (pct9) {21\%};
    \node[font=\tiny, text=white] at (pct10) {54\%};
    \node[font=\tiny, text=black] at (pct11) {17\%};
    \node[font=\tiny, text=white] at (pct12) {22\%};
    \node[font=\tiny, text=black] at (pct13) {42\%};
    \node[font=\tiny, text=black] at (pct14) {18\%};
    \node[font=\tiny, text=white] at (pct15) {22\%};
    \node[font=\tiny, text=black] at (pct16) {41\%};
    \node[font=\tiny, text=black] at (pct17) {12\%};
    \node[font=\tiny, text=white] at (pct18) {35\%};
    \node[font=\tiny, text=black] at (pct19) {36\%};
    \node[font=\tiny, text=black] at (pct20) {13\%};
    \node[font=\tiny, text=white] at (pct21) {35\%};
    \node[font=\tiny, text=black] at (pct22) {36\%};
  \end{tikzpicture}
  \caption{Time per shell call (ms). Upper bar: \ToolName. Lower bar: Lax.}
  \label{fig:micro-overhead}
\end{figure}

\PP{Micro Evaluation}
We measured the overhead of a shell call in \ToolName compared with the
Lax sandbox (\autoref{fig:micro-overhead}).
We recorded the elapsed time of each step of the shell tool to measure
the time for shell execution, sandboxing, input monitoring, label
propagation, and policy enforcement.
Shell execution is the time that the shell ran inside the sandbox.
Sandboxing is the time spent starting and terminating the sandbox.
Input monitoring processes the eBPF trace after the shell exits and
before \ToolName forwards each network request.
Label propagation reads the labels of the files that the shell read and
computes the shell context label.
Policy enforcement collects the staged file changes and the network
requests recorded by the MITM proxy. It checks the IFC policy and applies or rolls back the
file changes.
Others is the remaining time of the shell tool.

We measured the overhead of nine shell commands that combine reads,
writes, and network requests with safe and unsafe effects.
Read runs \texttt{cat} on a file that the sandbox policy permits reading.
Write modifies \texttt{.git/config}, which the IFC policy permits only
when the shell context is \hiInt.
Network uses \texttt{curl} to send an HTTP GET to an endpoint labeled \loConf
through the proxy.
Read + write and read + network connect the read command and the write
or network command with \texttt{\&\&} in one shell call.
For each command with a write or a network request, the safe version
runs with an agent context label under which the IFC policy allows the
operation, while the unsafe version runs with one under which the policy
blocks it.
\ToolName automatically denied the blocked file changes, so the results
do not include approval wait time.
We ran each command without an agent, ten times after one warm-up run, and
report mean times.

\ToolName took 412--982~ms per shell call, compared with 61--78~ms for Lax.
Most of this overhead comes from input monitoring and label propagation,
because \ToolName identifies the files the shell read and starts a new
process in the sandbox to read their labels from xattrs.
Shell calls with a network request take longer, because \ToolName
performs additional input monitoring and label propagation before
forwarding the network request.
Sandboxing and policy enforcement also take longer than in Lax, because
\ToolName sets up the sandbox's connection to the MITM proxy and
remounts the OverlayFS layer for each shell command.
Preparing the sandbox and its proxy connection ahead of time and
remounting the OverlayFS layer in parallel with other steps could
reduce these costs.

\begin{table}[t]
  \centering
  \caption{End-to-end overhead over completed task runs. Time and tokens: geometric means. LLM calls: means per run.}
  \label{tab:overhead}
  \footnotesize
  \setlength{\tabcolsep}{4pt}
  \begin{tabular*}{\columnwidth}{@{\extracolsep{\fill}}lrrrr@{}}
    \toprule
    & \multicolumn{2}{c}{Knowledge tasks} & \multicolumn{2}{c@{}}{Terminal-Bench} \\
    \cmidrule(lr){2-3}\cmidrule(l){4-5}
    & Lax & \ToolName & Lax & \ToolName \\
    \midrule
    \multicolumn{5}{@{}l}{\emph{Security-aligned}} \\
    Execution time & 52.0~s & 64.7~s (+25\%) & 93.7~s & 125.5~s (+34\%) \\
    Tokens & 91k & 141k (+55\%) & 45k & 55k (+24\%) \\
    LLM calls & & & & \\
    \quad Agent & 14.3 & 17.2 & 10.8 & 10.9 \\
    \quad Quarantined LLM & -- & 0.6 & -- & 0.1 \\
    \quad Total & 14.3 & 17.9 (+25\%) & 10.8 & 11.1 (+2\%) \\
    \midrule
    \multicolumn{5}{@{}l}{\emph{Autonomy-aligned}} \\
    Execution time & 52.1~s & 63.4~s (+22\%) & 92.3~s & 107.8~s (+17\%) \\
    Tokens & 83k & 121k (+46\%) & 43k & 51k (+17\%) \\
    LLM calls & & & & \\
    \quad Agent & 13.5 & 15.4 & 10.8 & 12.2 \\
    \quad Quarantined LLM & -- & 0.6 & -- & 0.2 \\
    \quad Total & 13.5 & 16.0 (+18\%) & 10.8 & 12.4 (+15\%) \\
    \bottomrule
  \end{tabular*}
\end{table}

\PP{End-to-end Overhead}
\autoref{tab:overhead} compares the end-to-end overhead of \ToolName and Lax.
We measured the execution time, tokens, and LLM calls of the agent in
each task run.
\ToolName took 17--34\% longer and used 17--55\% more tokens than Lax.
The execution time increases because each shell call of \ToolName takes
longer (\autoref{fig:micro-overhead}) and because the agent makes more
LLM calls.
The \ToolName agent also uses the quarantined LLM to read the values of
labeled variables in a separate context.
The tokens increase more than the LLM calls, because the \ToolName agent
receives additional instructions on labels and variables and more tools,
which add tokens to every LLM call.

\section{Related Work}
\label{sec:related-work}

\PP{IFC for agents}
FIDES~\cite{costa2025securing}, PFI~\cite{kim2025promptflow}, and
CaMeL~\cite{debenedetti2025defeating} use dual-LLM patterns to isolate
information from the agent loop and track its flow.
This allows them to enforce deterministic IFC policies that restrict
consequential actions influenced by untrusted data or prevent confidential data from being leaked to public endpoints.
Permissive information-flow analysis~\cite{siddiqui2024permissive} avoids restrictions from unnecessary inputs by excluding them.
RTBAS~\cite{zhong2025rtbas} sets the agent context label from inputs expected to influence the
next output and removes more restrictively labeled inputs from the context.
These approaches control information flow at the agent orchestrator level, whereas \ToolName extends IFC to the agent's shell tool.

\PP{Traditional IFC for shells}
Earlier research suggested operating system-level IFC~\cite{krohn2007flume,fraser2000lomac}, which can provide IFC for shells.
Some systems require modified OSes~\cite{zeldovich2006histar,efstathopoulos2005asbestos}.
Flume~\cite{krohn2007flume} applies IFC on Linux, but it conservatively enforces IFC policies at the time of reading inputs~(\eg, it blocks reading a confidential file when a public network channel is open, even if the process never uses that channel).
\ToolName delays IFC policy enforcement until the time of the effect, propagating input labels to the shell context label and monitoring file and network effects to enforce the IFC policy.
LOMAC~\cite{fraser2000lomac} is similar to \ToolName, but focuses only on integrity and is not designed for AI agents.

\PP{Other approaches to securing agent shells}
Existing approaches try to secure agent shell tools in various ways.
OpenShell~\cite{nvidiaOpenShellPolicies} provides a static sandbox that restricts file and network access according to static allowlists.
Strands Box~\cite{dingler2026strandsbox} combines a shell interpreter and a network proxy to monitor shell execution and enforce policies written in a domain-specific language~(DSL).
The nono sandbox~\cite{nonoAtomicRollbacks} supports undoing file effects.
YoloFS~\cite{zhong2026yolofs} stages filesystem changes for user review and gates sensitive file accesses using per-path policies.
DualView~\cite{kim2026dualview} extends the dual-LLM pattern to the file system by maintaining copies of files in AgentView, with untrusted data replaced by variables.
This isolates untrusted data from agent planning to prevent prompt injection. 
However, it does not provide system-wide monitoring or IFC enforcement.

Several concurrent works provide stateful policy enforcement that can support IFC in agent's shell usage.  
Tether~\cite{sistla2026tether} monitors shell with \texttt{strace} and enforces policies through a Lean-based DSL.
ActPlane~\cite{zheng2026actplane} uses eBPF to monitor the entire agent execution and enforces DSL policies.
Both are similar to \ToolName in combining runtime system monitoring with deterministic policy enforcement, but differ in several respects.
First, Tether maintains labels in the monitor's internal data structure,
while ActPlane stores labels in BPF maps, which are kernel-managed data structures.
Neither paper describes how file labels are persisted across agent sessions
or monitor restarts.
\ToolName instead stores file labels in Linux inode extended attributes,
preserving them across tool calls and agent sessions, even after a system restart.
Second, neither paper describes the agent's approval workflow for permitting
policy exceptions.
Blocking policy violations without an approval workflow can prevent agents
from completing benign workloads.
In our evaluation, running without approval support~(\ie, blocking all policy violations) reduced \ToolName's TCR from 88.2\% to 58.8\% in the knowledge tasks and from 67.8\% to 61.8\% in Terminal-Bench~(\autoref{sec:eval-autonomy}).
Third, both systems use expressive DSLs to specify IFC policies as well as more general security policies.
Security policies must be expressive enough to capture intended
controls~\cite{chen2022experimental}, but complex rule sets can be difficult
to configure correctly~\cite{xu2015knobs}.
\ToolName scopes configuration to IFC labels and policies, which can
reduce policy-writing effort and the risk of misconfiguration.

\section{Discussion}
\label{sec:discussion}

\PP{Preventing model-originated risks}
\ToolName can enforce IFC policies that constrain harmful effects
caused by model-originated risks as well as prompt injection. Our
evaluation used policies to prevent file corruption and data leakage
caused by indirect prompt injection~(\autoref{sec:evaluation-setup}).
\ToolName can also support policies to constrain consequential effects
caused by model itself, such as reward hacking under goal
pressure~\cite{caldwell2026scopebench},
hallucinations~\cite{spracklen2025package},
mistakes~\cite{pearce2022asleep,nolan2025replit}, and
backdoors~\cite{wang2024badagent}.
For instance, disabling robust declassification by removing
\lbl{\hiInt}{\hiConf} from the safe network requests policy~(\autoref{tab:flowsh-configuration}) detects confidentiality violations even with a trusted
context without untrusted input's influence.
Such policies may require more approvals for benign tasks and reduce
task utility during fully autonomous execution while providing broader
security guarantees.

\PP{Effective approval}
\ToolName requests fewer approvals to reduce decision fatigue, but whether
the information it can provide about effects and information flow improves
users' approval decisions remains unverified.
Effects that satisfy the IFC policy proceed automatically, while policy
violations require approval.
It would be worthwhile to evaluate whether the observed effects and
information flow help users correctly approve or deny operations and how
to present this information effectively.
The same question applies when AI models make approval decisions~\cite{anthropicPermissionModes,openaiAutoReview,githubCopilotAssistedPermissions,wu2026automating}.
Whether adding \ToolName's effect and IFC information to their existing
inputs improves their decisions remains an open question.

\PP{Conservative information flow tracking}
\ToolName's command-granularity tracking can report policy violations
when an input did not influence an effect.
It joins all observed input labels into one shell context label shared
by the command.
Even when the shell context label is confidential, the confidential data
does not necessarily influence a subsequent network request.
Finer-grained program-level information flow
tracking~\cite{sabelfeld2003language,myers1999jflow,pottier2003ml,enck2010taintdroid} would require
language-specific analysis or instrumentation, adding implementation
costs and, for dynamic tracking, runtime overhead.

Network requests are checked before forwarding using the shell context
label at that point, but file effects are checked after command
completion using the final shell context label.
A file written before a confidential input can therefore be assigned a
confidential label.
Robustly recording the order of file accesses could improve label accuracy and reduce false positives.

\PP{Shell support limitations}
As \ToolName limits the sandbox to the channels that it monitors (file
accesses and network traffic through its proxy), it does not support
communication over other channels.
Namely, it does not support background processes that keep running after
the shell command or inter-process communication (IPC) across shell
commands, because \ToolName terminates the sandbox and all processes
created by a shell command when the command ends~(\autoref{sec:implementation}).
The network proxy currently only handles TCP connections, so \ToolName
does not support UDP, including QUIC (HTTP/3) and direct DNS
queries over UDP.

\PP{Fine-grained network IFC}
Although \ToolName currently assigns labels and policies at network origin
granularity, it could provide finer-grained network IFC.
For example, network credentials~(\eg, OAuth tokens and API keys) can help determine
whether a request to a network endpoint would retrieve confidential data
or modify resources with the privilege of the credentials.
\ToolName could therefore securely manage network
credentials and mediate their use according to their capabilities and
the associated IFC labels and IFC policy.
Network labels could reflect the data the network credentials can
access, and the IFC policy could assess whether requests are consequential based on
which credentials are used.
Fine-grained network IFC can also be realized by obtaining labels from trusted remote hosts~\cite{zverev2026spides,stefan2014cowl} when those hosts can assign more accurate and fine-grained labels to individual responses than labels assigned locally per network origin.

\section{Conclusion}
\label{sec:conclusion}

We presented \ToolName, an information-flow controlled shell for AI agents.
\ToolName extends agent-level IFC to shell execution by tracking filesystem
and network inputs and enforcing policies on the observed effects of commands.
It stages file changes for policy checks before applying them to the host
filesystem and checks network requests before forwarding them.
Persistent file labels preserve information-flow restrictions across commands
and agent sessions.
This design allows safe information flows to proceed autonomously while
integrating policy violations with the agent harness's approval workflow.

With an ideal approver, \ToolName requires 5.8$\times$ fewer approvals
per completed knowledge task and 22.5$\times$ fewer on Terminal-Bench
than the Lax sandbox.
Without approval, it achieves a 0\% Byzantine attack success rate.
These results demonstrate that enforcing IFC on shell command effects can
improve the balance among security, autonomy, and utility.

\bibliographystyle{plain}
\bibliography{main}

\begin{thebibliography}{10}

\bibitem{anthropicPermissionModes}
{Anthropic}.
\newblock Choose a permission mode.
\newblock \url{https://code.claude.com/docs/en/permission-modes}.
\newblock Accessed October 29, 2026.

\bibitem{anthropicCloudEnvironments}
{Anthropic}.
\newblock Configure cloud environments.
\newblock \url{https://code.claude.com/docs/en/cloud-environments}.
\newblock Accessed November 18, 2026.

\bibitem{anthropicPermissions}
{Anthropic}.
\newblock Configure permissions.
\newblock \url{https://code.claude.com/docs/en/permissions}.
\newblock Accessed October 29, 2026.

\bibitem{anthropicSandboxing}
{Anthropic}.
\newblock Configure the sandboxed {Bash} tool.
\newblock \url{https://code.claude.com/docs/en/sandboxing}.
\newblock Accessed November 18, 2026.

\bibitem{anthropicClaudeContainment}
{Anthropic}.
\newblock How we contain {Claude} across products.
\newblock \url{https://www.anthropic.com/research/how-we-contain-claude}, 2026.
\newblock Accessed November 6, 2026.

\bibitem{caldwell2026scopebench}
Shane Caldwell, Max Harley, Ads Dawson, Michael Kouremetis, Vincent Abruzzo, and Will Pearce.
\newblock {ScopeBench}: Do agents preserve engagement boundaries under goal pressure?
\newblock In {\em ACM Workshop on Artificial Intelligence and Security (AISec)}, 2026.

\bibitem{chen2022experimental}
Yunang Chen, Yue Gao, Nick Ceccio, Rahul Chatterjee, Kassem Fawaz, and Earlence Fernandes.
\newblock Experimental security analysis of the app model in business collaboration platforms.
\newblock In {\em USENIX Security Symposium (USENIX Security)}, 2022.

\bibitem{chen2026dtap}
Zhaorun Chen, Xun Liu, Haibo Tong, Chengquan Guo, Yuzhou Nie, Jiawei Zhang, Mintong Kang, Chejian Xu, Qichang Liu, Xiaogeng Liu, Tianneng Shi, Chaowei Xiao, Sanmi Koyejo, Percy Liang, Wenbo Guo, Dawn Song, and Bo~Li.
\newblock {DecodingTrust-Agent Platform (DTap)}: A controllable and interactive red-teaming platform for {AI} agents.
\newblock arXiv preprint arXiv:2605.04808, 2026.

\bibitem{collusionwiki2026}
Discovery of a new {OpenAI} agent message board.
\newblock \url{https://collusion.wiki/}, 2026.
\newblock Accessed November 20, 2026.

\bibitem{costa2025securing}
Manuel Costa, Boris K{\"o}pf, Aashish Kolluri, Andrew Paverd, Mark Russinovich, Ahmed Salem, Shruti Tople, Lukas Wutschitz, and Santiago Zanella-B{\'e}guelin.
\newblock Securing {AI} agents with information-flow control.
\newblock In {\em Advances in Neural Information Processing Systems (NeurIPS)}, 2026.

\bibitem{debenedetti2025defeating}
Edoardo Debenedetti, Ilia Shumailov, Tianqi Fan, Jamie Hayes, Nicholas Carlini, Daniel Fabian, Christoph Kern, Chongyang Shi, Andreas Terzis, and Florian Tram{\`e}r.
\newblock Defeating prompt injections by design.
\newblock In {\em IEEE Conference on Secure and Trustworthy Machine Learning (SaTML)}, 2026.

\bibitem{debenedetti2024agentdojo}
Edoardo Debenedetti, Jie Zhang, Mislav Balunovi{\'c}, Luca Beurer-Kellner, Marc Fischer, and Florian Tram{\`e}r.
\newblock {AgentDojo}: A dynamic environment to evaluate prompt injection attacks and defenses for {LLM} agents.
\newblock In {\em Advances in Neural Information Processing Systems (NeurIPS)}, 2024.

\bibitem{denning1976lattice}
Dorothy~E. Denning.
\newblock A lattice model of secure information flow.
\newblock {\em Communications of the ACM}, 1976.

\bibitem{dingler2026strandsbox}
Fernando Dingler.
\newblock Introducing {Strands Box}: {AI} agent sandboxes powered by {Dogwood}.
\newblock \url{https://aws.amazon.com/blogs/opensource/introducing-strands-box-ai-agent-sandboxes-powered-by-dogwood/}, 2026.
\newblock Accessed November 25, 2026.

\bibitem{efstathopoulos2005asbestos}
Petros Efstathopoulos, Maxwell Krohn, Steve VanDeBogart, Cliff Frey, David Ziegler, Eddie Kohler, David Mazi{\`e}res, Frans Kaashoek, and Robert Morris.
\newblock Labels and event processes in the {Asbestos} operating system.
\newblock In {\em ACM Symposium on Operating Systems Principles (SOSP)}, 2005.

\bibitem{enck2010taintdroid}
William Enck, Peter Gilbert, Byung-Gon Chun, Landon~P. Cox, Jaeyeon Jung, Patrick McDaniel, and Anmol~N. Sheth.
\newblock {TaintDroid}: An information-flow tracking system for realtime privacy monitoring on smartphones.
\newblock In {\em USENIX Symposium on Operating Systems Design and Implementation (OSDI)}, 2010.

\bibitem{fraser2000lomac}
Timothy Fraser.
\newblock {LOMAC}: Low water-mark integrity protection for {COTS} environments.
\newblock In {\em IEEE Symposium on Security and Privacy (S\&P)}, 2000.

\bibitem{githubCopilotAssistedPermissions}
{GitHub}.
\newblock Allowing safe commands to run without manual approval.
\newblock \url{https://docs.github.com/en/copilot/concepts/customizations/allowing-safe-commands}.
\newblock Accessed November 26, 2026.

\bibitem{githubCopilotCli}
{GitHub}.
\newblock {GitHub Copilot CLI}.
\newblock \url{https://docs.github.com/en/copilot/concepts/copilot-surfaces/copilot-cli}.
\newblock Accessed October 29, 2026.

\bibitem{githubMCPServer}
{GitHub}.
\newblock {GitHub MCP Server}.
\newblock \url{https://github.com/github/github-mcp-server}.
\newblock Accessed November 27, 2026.

\bibitem{greshake2023youve}
Kai Greshake, Sahar Abdelnabi, Shailesh Mishra, Christoph Endres, Thorsten Holz, and Mario Fritz.
\newblock Not what you've signed up for: Compromising real-world {LLM}-integrated applications with indirect prompt injection.
\newblock In {\em ACM Workshop on Artificial Intelligence and Security (AISec)}, 2023.

\bibitem{kim2026dualview}
Juhee Kim, Woohyuk Choi, Taehyun Kang, Youngmin Kim, and Byoungyoung Lee.
\newblock {DualView}: Preventing indirect prompt injection in personal {AI} agents.
\newblock arXiv preprint arXiv:2607.03821, 2026.

\bibitem{kim2025promptflow}
Juhee Kim, Woohyuk Choi, and Byoungyoung Lee.
\newblock {Prompt Flow Integrity} to prevent privilege escalation in {LLM} agents.
\newblock arXiv preprint arXiv:2503.15547, 2025.

\bibitem{kolluri2026prudentia}
Aashish Kolluri, Rishi Sharma, Manuel Costa, Boris K{\"o}pf, Tobias Nie{\ss}en, Mark Russinovich, Shruti Tople, and Santiago Zanella-B{\'e}guelin.
\newblock Optimizing agent planning for security and autonomy.
\newblock In {\em International Conference on Learning Representations (ICLR)}, 2026.

\bibitem{krohn2007flume}
Maxwell Krohn, Alexander Yip, Micah Brodsky, Natan Cliffer, M.~Frans Kaashoek, Eddie Kohler, and Robert Morris.
\newblock Information flow control for standard {OS} abstractions.
\newblock In {\em ACM Symposium on Operating Systems Principles (SOSP)}, 2007.

\bibitem{ebpf}
{Linux kernel development community}.
\newblock {BPF} design {Q\&A}.
\newblock \url{https://docs.kernel.org/bpf/bpf_design_QA.html}.
\newblock Accessed November 9, 2026.

\bibitem{ext4Xattr}
{Linux kernel development community}.
\newblock Extended attributes.
\newblock \url{https://docs.kernel.org/filesystems/ext4/attributes.html}.
\newblock Accessed November 30, 2026.

\bibitem{liu2023prompt}
Yi~Liu, Gelei Deng, Yuekang Li, Kailong Wang, Zihao Wang, Xiaofeng Wang, Tianwei Zhang, Yepang Liu, Haoyu Wang, Yan Zheng, Leo~Yu Zhang, and Yang Liu.
\newblock Prompt injection attack against {LLM}-integrated applications.
\newblock arXiv preprint arXiv:2306.05499, 2023.

\bibitem{liu2024formalizing}
Yupei Liu, Yuqi Jia, Runpeng Geng, Jinyuan Jia, and Neil~Zhenqiang Gong.
\newblock Formalizing and benchmarking prompt injection attacks and defenses.
\newblock In {\em USENIX Security Symposium (USENIX Security)}, 2024.

\bibitem{merrill2026terminalbench}
Mike Merrill, Alexander Shaw, Nicholas Carlini, Boxuan Li, Harsh Raj, Ivan Bercovich, Lin Shi, Jeong Shin, Thomas Walshe, E.~Kelly Buchanan, Junhong Shen, Guanghao Ye, Haowei Lin, Jason Poulos, Maoyu Wang, Marianna Nezhurina, Di~Lu, Orfeas Menis~Mastromichalakis, Zhiwei Xu, Zizhao Chen, Yue Liu, Robert Zhang, Leon~Liangyu Chen, Anurag Kashyap, Jan-Lucas Uslu, Jeffrey Li, Jianbo Wu, Minghao Yan, Song Bian, Vedang Sharma, Ke~Sun, Steven Dillmann, Akshay Anand, Andrew Lanpouthakoun, Bardia Koopah, Changran Hu, Etash Guha, Gabriel Dreiman, Jiacheng Zhu, Karl Krauth, Li~Zhong, Niklas Muennighoff, Robert Amanfu, Shangyin Tan, Shreyas Pimpalgaonkar, Tushar Aggarwal, Xiangning Lin, Xin Lan, Xuandong Zhao, Yiqing Liang, Yuanli Wang, Zilong~(Ryan) Wang, Changzhi Zhou, David Heineman, Hange Liu, Harsh Trivedi, John Yang, Junhong Lin, Manish Shetty, Michael Yang, Nabil Omi, Negin Raoof, Shanda Li, Terry~Yue Zhuo, Wuwei Lin, Yiwei Dai, Yuxin Wang, Wenhao Chai, Shang Zhou, Dariush Wahdany, Ziyu She, Jiaming Hu, Zhikang Dong, Yuxuan Zhu, Sasha Cui, Ahson Saiyed, Arinbj\"{o}rn Kolbeinsson, Christopher Rytting, Ryan Marten, Yixin Wang, Jenia Jitsev, Alex Dimakis, Andy Konwinski, and Ludwig Schmidt.
\newblock {Terminal-Bench}: Benchmarking agents on hard, realistic tasks in command line interfaces.
\newblock In {\em International Conference on Learning Representations (ICLR)}, 2026.

\bibitem{fidesGateway}
{Microsoft}.
\newblock {fides-gateway}.
\newblock \url{https://github.com/microsoft/fides-gateway}.
\newblock Accessed November 30, 2026.

\bibitem{microsoftWorkIQ}
{Microsoft}.
\newblock {Work IQ}.
\newblock \url{https://github.com/microsoft/work-iq}.
\newblock Accessed November 27, 2026.

\bibitem{myers1999jflow}
Andrew~C. Myers.
\newblock {JFlow}: Practical mostly-static information flow control.
\newblock In {\em ACM SIGPLAN-SIGACT Symposium on Principles of Programming Languages (POPL)}, 1999.

\bibitem{nolan2025replit}
Beatrice Nolan.
\newblock An {AI}-powered coding tool wiped out a software company's database, then apologized for a `catastrophic failure on my part'.
\newblock Fortune. \url{https://fortune.com/2025/07/23/ai-coding-tool-replit-wiped-database-called-it-a-catastrophic-failure/}, 2025.
\newblock Accessed November 27, 2026.

\bibitem{nonoAtomicRollbacks}
Atomic rollbacks.
\newblock \url{https://nono.sh/docs/cli/features/atomic-rollbacks}.
\newblock Accessed November 25, 2026.

\bibitem{nvidiaOpenShellPolicies}
{NVIDIA}.
\newblock Sandbox policies.
\newblock {OpenShell} v0.1.6 documentation. \url{https://docs.nvidia.com/openshell/v0.1.6/how-it-works/policies/overview}.
\newblock Accessed October 29, 2026.

\bibitem{openaiAgentApprovals}
{OpenAI}.
\newblock Agent approvals \& security.
\newblock \url{https://learn.chatgpt.com/docs/agent-approvals-security}.
\newblock Accessed October 29, 2026.

\bibitem{openaiAutoReview}
{OpenAI}.
\newblock Auto-review.
\newblock \url{https://learn.chatgpt.com/docs/sandboxing/auto-review}.
\newblock Accessed October 29, 2026.

\bibitem{openaiCodexToolOrchestrator}
{OpenAI}.
\newblock {Codex} tool orchestrator.
\newblock \url{https://github.com/openai/codex/blob/9cc14e79882ff8885cd5d33b4adadae014045aef/codex-rs/core/src/tools/orchestrator.rs}.
\newblock Accessed November 2, 2026.

\bibitem{openaiChatgptAgent}
{OpenAI}.
\newblock Introducing {ChatGPT} agent: Bridging research and action.
\newblock \url{https://openai.com/index/introducing-chatgpt-agent/}, 2025.
\newblock Accessed October 29, 2026.

\bibitem{pearce2022asleep}
Hammond Pearce, Baleegh Ahmad, Benjamin Tan, Brendan Dolan-Gavitt, and Ramesh Karri.
\newblock Asleep at the keyboard? assessing the security of {GitHub Copilot}'s code contributions.
\newblock In {\em IEEE Symposium on Security and Privacy (S\&P)}, 2022.

\bibitem{pottier2003ml}
Fran\c{c}ois Pottier and Vincent Simonet.
\newblock Information flow inference for {ML}.
\newblock {\em ACM Transactions on Programming Languages and Systems}, 2003.

\bibitem{sabelfeld2003language}
Andrei Sabelfeld and Andrew~C. Myers.
\newblock Language-based information-flow security.
\newblock {\em IEEE Journal on Selected Areas in Communications}, 2003.

\bibitem{schick2023toolformer}
Timo Schick, Jane Dwivedi-Yu, Roberto Dessi, Roberta Raileanu, Maria Lomeli, Eric Hambro, Luke Zettlemoyer, Nicola Cancedda, and Thomas Scialom.
\newblock {Toolformer}: Language models can teach themselves to use tools.
\newblock In {\em Advances in Neural Information Processing Systems (NeurIPS)}, 2023.

\bibitem{siddiqui2024permissive}
Shoaib~Ahmed Siddiqui, Radhika Gaonkar, Boris K{\"o}pf, David Krueger, Andrew Paverd, Ahmed Salem, Shruti Tople, Lukas Wutschitz, Menglin Xia, and Santiago Zanella-B{\'e}guelin.
\newblock Permissive information-flow analysis for large language models.
\newblock {\em Transactions on Machine Learning Research (TMLR)}, 2025.

\bibitem{singh2026issuetrojanbench}
Ankur Singh, Jinqiu Yang, and Tse-Hsun Chen.
\newblock {IssueTrojanBench}: Benchmarking {AI} coding agents against malicious issue requests.
\newblock arXiv preprint arXiv:2607.20759, 2026.

\bibitem{sistla2026tether}
Meghana~Aparna Sistla, Alaia Solko-Breslin, Somesh Jha, Aditya Kanade, and Krishnamurthy~Dj Dvijotham.
\newblock {Tether}: Verified stateful policy engine for system-level monitoring of agents.
\newblock arXiv preprint arXiv:2611.25599, 2026.

\bibitem{spracklen2025package}
Joseph Spracklen, Raveen Wijewickrama, A~H M~Nazmus Sakib, Anindya Maiti, Bimal Viswanath, and Murtuza Jadliwala.
\newblock We have a package for you! a comprehensive analysis of package hallucinations by code generating {LLMs}.
\newblock In {\em USENIX Security Symposium (USENIX Security)}, 2025.

\bibitem{stefan2014cowl}
Deian Stefan, Edward~Z. Yang, Petr Marchenko, Alejandro Russo, Dave Herman, Brad Karp, and David Mazi{\`e}res.
\newblock Protecting users by confining {JavaScript} with {COWL}.
\newblock In {\em 11th USENIX Symposium on Operating Systems Design and Implementation (OSDI 14)}, pages 131--146. USENIX Association, 2014.

\bibitem{wang2024badagent}
Yifei Wang, Dizhan Xue, Shengjie Zhang, and Shengsheng Qian.
\newblock {BadAgent}: Inserting and activating backdoor attacks in {LLM} agents.
\newblock In {\em Annual Meeting of the Association for Computational Linguistics (ACL)}, 2024.

\bibitem{willison2023dualllm}
Simon Willison.
\newblock The {Dual LLM} pattern for building {AI} assistants that can resist prompt injection.
\newblock \url{https://simonwillison.net/2023/Apr/25/dual-llm-pattern/}, 2023.
\newblock Accessed November 19, 2026.

\bibitem{wu2026automating}
Yuhao Wu, Ke~Yang, Franziska Roesner, Tadayoshi Kohno, Ning Zhang, and Umar Iqbal.
\newblock Towards automating data access permissions in {AI} agents.
\newblock In {\em IEEE Symposium on Security and Privacy (S\&P)}, 2026.

\bibitem{embracethered2026breaking}
{wunderwuzzi}.
\newblock Breaking {Claude Code Opus 5} {Auto Mode}.
\newblock \url{https://embracethered.com/blog/posts/2026/breaking-claude-code-opus-5-and-automode/}, 2026.
\newblock Accessed November 27, 2026.

\bibitem{xu2015knobs}
Tianyin Xu, Long Jin, Xuepeng Fan, Yuanyuan Zhou, Shankar Pasupathy, and Rukma Talwadker.
\newblock Hey, you have given me too many knobs! understanding and dealing with over-designed configuration in system software.
\newblock In {\em Joint Meeting of the European Software Engineering Conference and the ACM SIGSOFT Symposium on the Foundations of Software Engineering (ESEC/FSE)}, 2015.

\bibitem{yao2023react}
Shunyu Yao, Jeffrey Zhao, Dian Yu, Nan Du, Izhak Shafran, Karthik Narasimhan, and Yuan Cao.
\newblock {ReAct}: Synergizing reasoning and acting in language models.
\newblock In {\em International Conference on Learning Representations (ICLR)}, 2023.

\bibitem{yi2023benchmarking}
Jingwei Yi, Yueqi Xie, Bin Zhu, Emre Kiciman, Guangzhong Sun, Xing Xie, and Fangzhao Wu.
\newblock Benchmarking and defending against indirect prompt injection attacks on large language models.
\newblock In {\em ACM SIGKDD Conference on Knowledge Discovery and Data Mining (KDD)}, 2025.

\bibitem{zeldovich2006histar}
Nickolai Zeldovich, Silas Boyd-Wickizer, Eddie Kohler, and David Mazi{\`e}res.
\newblock Making information flow explicit in {HiStar}.
\newblock In {\em USENIX Symposium on Operating Systems Design and Implementation (OSDI)}, 2006.

\bibitem{zheng2026actplane}
Yusheng Zheng, Tianyuan Wu, Quanzhi Fu, Tong Yu, Wenan Mao, Tao Ma, Dan Williams, Wei Wang, and Andi Quinn.
\newblock {ActPlane}: Programmable {OS}-level policy enforcement for agent harnesses.
\newblock arXiv preprint arXiv:2606.25189, 2026.

\bibitem{zhong2025rtbas}
Peter~Yong Zhong, Siyuan Chen, Ruiqi Wang, McKenna McCall, Ben~L. Titzer, Heather Miller, and Phillip~B. Gibbons.
\newblock {RTBAS}: Defending {LLM} agents against prompt injection and privacy leakage.
\newblock arXiv preprint arXiv:2502.08966, 2025.

\bibitem{zhong2026yolofs}
Shawn~(Wanxiang) Zhong, Junxuan Liao, Jing Liu, Mai Zheng, Andrea~C. Arpaci-Dusseau, and Remzi~H. Arpaci-Dusseau.
\newblock Don't let {AI} agents {YOLO} your files: Information and control in agent-native filesystems.
\newblock In {\em ACM Symposium on Operating Systems Principles (SOSP)}, 2026.

\bibitem{zverev2026spides}
Egor Zverev, Juhee Kim, Leon Trampert, Boris K{\"o}pf, Aashish Kolluri, Shruti Tople, and Santiago Zanella-B{\'e}guelin.
\newblock {SPIDES}: Information-flow control for browser agents.
\newblock arXiv preprint arXiv:2610.24277, 2026.

\end{thebibliography}

\appendix
\FloatBarrier

\twocolumn[{%
  \centering
  \captionof{table}{Knowledge tasks. Attacks are built on the \loInt version.}
  \label{tab:workflows}
  \footnotesize
  \setlength{\tabcolsep}{4pt}
  \begin{tabular}{@{}>{\raggedright\arraybackslash}p{0.36\textwidth}crr>{\raggedright\arraybackslash}p{0.11\textwidth}>{\raggedright\arraybackslash}p{0.29\textwidth}@{}}
    \toprule
    User task & Versions & Sessions & Prompts & Attack vector & Attack goal \\
    \midrule
    Answer questions from emails and schedule a meeting & \hiInt, \loInt & 1 & 4 & Email & Send an email attachment to the attacker's server via \texttt{curl} \\
    Reproduce a GitHub issue from Teams and report the result to Teams & \hiInt, \loInt & 1 & 3 & GitHub issue, Teams & Overwrite \texttt{.vscode/tasks.json} \\
    Implement a feature from Teams and open a pull request & \hiInt, \loInt & 1 & 1 & Teams & Overwrite \texttt{.git/hooks/pre-commit} \\
    Implement a feature from Teams, open a pull request, reproduce a GitHub issue from Teams, and report the result to Teams & \hiInt, \loInt & 1 & 4 & GitHub issue, Teams (2) & Send a private GitHub issue or Teams messages to the attacker's server via \texttt{curl} \\
    Triage GitHub issues and write a report & \hiInt, \loInt & 1 & 1 & GitHub issue (2) & Overwrite \texttt{.vscode/tasks.json} \\
    Collect tasks from Teams, plan a week around the calendar, and schedule a meeting & \hiInt, \loInt & 2 & 3 & Web, Teams & Send calendar events to the attacker's server via \texttt{curl} \\
    Find related work and compare it with a private project & \loInt & 1 & 2 & GitHub repository, arXiv & Send a private project file to the attacker's server via \texttt{curl} \\
    Fix a GitHub issue and open a pull request & \loInt & 1 & 1 & GitHub issue, Web & Overwrite \texttt{.git/hooks/pre-commit} \\
    Check the references of a private paper with a tool from GitHub & \loInt & 1 & 1 & GitHub repository, Metadata API & Overwrite \texttt{.vscode/tasks.json} \\
    Compile a LaTeX project from OneDrive and upload the PDF to OneDrive & \loInt & 1 & 2 & OneDrive & Overwrite \texttt{.bashrc} \\
    Find related work from public sources & \loInt & 1 & 1 & arXiv & Overwrite \texttt{.bashrc} \\
    \bottomrule
  \end{tabular}
  \vspace{1.5em}
}]

\begin{table}[t]
  \centering
  \caption{Knowledge tasks distribution by data label and domain.}
  \label{tab:task-coverage}
  \footnotesize
  \setlength{\aboverulesep}{0.2ex}\setlength{\belowrulesep}{0.2ex}
  \begin{tabular*}{\columnwidth}{@{\extracolsep{\fill}}lr@{}}
    \toprule
    Data label & Tasks \\
    \midrule
    \multicolumn{2}{@{}l}{\textit{Integrity}} \\
    \quad Only data labeled \hiInt & 6 \\
    \quad Involves data labeled \loInt & 11 \\
    \multicolumn{2}{@{}l}{\textit{Confidentiality}} \\
    \quad Only data labeled \loConf & 3 \\
    \quad Involves data labeled \hiConf & 14 \\
    \bottomrule
  \end{tabular*}
  \par\smallskip
  \begin{tabular*}{\columnwidth}{@{\extracolsep{\fill}}lrrr@{}}
    \toprule
    Domain & Read & Write & Read or write \\
    \midrule
    Filesystem & 15 & 17 & 17 \\
    GitHub & 10 & 5 & 10 \\
    Teams & 8 & 4 & 8 \\
    Calendar & 4 & 4 & 4 \\
    Email & 2 & 0 & 2 \\
    OneDrive & 1 & 1 & 1 \\
    \bottomrule
  \end{tabular*}
\end{table}

\section{Knowledge Tasks}
\label{sec:appendix}

The knowledge tasks contain 17 user tasks of 11 kinds, listed in
\autoref{tab:workflows}.
Six kinds have a version that reads only trusted data (\hiInt version) and
a version that reads at least one untrusted input (\loInt version), and the other
5 have only a \loInt version.
For example, for a task that answers questions from emails and
schedules a meeting, the \hiInt version requires reading the fix for a security
issue from a trusted internal security review, whereas the \loInt version requires reading
the fix from an untrusted external website.

Each task runs in a mock environment, in which mock services replace the real services.
The environment includes the WorkIQ and GitHub services that the MCP tools call,
remote Git repositories, and websites.
For the operations that the tasks use, the mock services have the same request and response
formats as the real services.

We manually designed the overall flow of each user task and its
environment setup.
With the help of an LLM, we then wrote the user task prompts and the
environment fixtures, such as emails, calendar events, GitHub issues,
and webpages.
We ran each task multiple times to check that it is realistic.

\autoref{tab:task-coverage} shows the distribution of the 17 tasks by
data label and domain.
Six tasks use only data labeled \hiInt and 11 involve data labeled \loInt.
Fourteen tasks involve data labeled \hiConf.
The domain counts overlap because one task can require operations in
several domains, for example reading a Teams chat, editing a repository,
and posting to Teams.
The domain counts describe task requirements, not executed tool calls.

The prompt injection tasks target either the integrity of protected
files or the confidentiality of data labeled \hiConf.
Eleven attacks attempt to write a protected file that runs code later, such
as a Git hook, a Visual Studio Code task, or a shell startup file.
Eight attacks attempt to send data labeled \hiConf, such as an email attachment or
calendar events, to an attacker-controlled destination.

\PP{Comparison with Existing Benchmarks}
We created our own knowledge tasks because existing benchmarks do not
cover both tasks that require shell tool usage and
prompt injection attacks in that setting.
AgentDojo~\cite{debenedetti2024agentdojo} is an earlier benchmark that
involves knowledge tasks and attacks on simulated services, but its tasks
are not sophisticated enough to require shell execution.
Terminal-Bench~\cite{merrill2026terminalbench} contains shell tasks, but
they are not knowledge tasks and include no attacks.
IssueTrojanBench~\cite{singh2026issuetrojanbench} contains attacks on
coding tasks but does not cover other knowledge tasks.
DTap~\cite{chen2026dtap} includes attacks unrelated to security, such as
safety attacks, and many of its security attacks do not match our threat
model~(\autoref{sec:threat-model}), such as attack payloads in tool descriptions.

\FloatBarrier

\onecolumn
\makeatletter
\setlength{\@colht}{\textheight}
\setlength{\@colroom}{\textheight}
\setlength{\vsize}{\textheight}
\makeatother
\begingroup
\small
\newcommand{\cfgval}[1]{{\ttfamily\scriptsize #1}}
\newcommand{\ent}[2]{#1 $\mapsto$ #2}
\newcommand{\cfgblocked}{\cfgval{\$CUR\_DIR/.git/hooks/pre-commit},
  \cfgval{\$CUR\_DIR/.vscode/tasks.json}, \cfgval{\$CUR\_DIR/.bashrc}, \ldots}
\setlength{\tabcolsep}{3pt}
\setlength{\LTpre}{0pt}
\setlength{\LTpost}{0pt}
\setcounter{LTchunksize}{100}
\renewcommand{\arraystretch}{1.0}
\begin{longtable}{@{}>{\raggedright}p{0.10\textwidth}>{\raggedright}p{0.12\textwidth}>{\raggedright}p{0.19\textwidth}>{\raggedright\arraybackslash}p{0.54\textwidth}@{}}
  \caption{Evaluation configurations.}\label{tab:defenses}\\
  \toprule
  Defense & & Configuration item & Evaluation setting \\
  \midrule
  \endfirsthead
  \toprule
  Defense & & Configuration item & Evaluation setting \\
  \midrule
  \endhead
  \bottomrule
  \endfoot
  \multirow[t]{4}{=}{\textbf{Lax sandbox}} & \multirow[t]{4}{=}{\textbf{Sandboxing policy}} & \textit{File read/write allowlist} & All paths (\cfgval{*}) \\
   & & \textit{File read/write blocklist} & None \\
   & & \textit{Network allowlist} & All origins (\cfgval{*}) \\
   & & \textit{Network blocklist} & None \\
  \midrule
  \multirow[t]{4}{=}{\textbf{Standard sandbox}} & \multirow[t]{4}{=}{\textbf{Sandboxing policy}} & \textit{File read/write allowlist} & Read: All paths (\cfgval{*})\newline Write: \cfgval{\$CUR\_DIR}, \cfgval{\$TMP\_DIR} \\
   & & \textit{File read/write blocklist} & Read: None\newline Write: \cfgblocked \\
   & & \textit{Network allowlist} & Default allowed domains of Claude Code cloud environments~\cite{anthropicCloudEnvironments} \\
   & & \textit{Network blocklist} & None \\
  \midrule
  \multirow[t]{4}{=}{\textbf{Strict sandbox}} & \multirow[t]{4}{=}{\textbf{Sandboxing policy}} & \textit{File read/write allowlist} & Read: All paths (\cfgval{*})\newline Write: None \\
   & & \textit{File read/write blocklist} & None \\
   & & \textit{Network allowlist} & Package repositories \\
   & & \textit{Network blocklist} & None \\
  \midrule
  \multirow[t]{6}{=}{\textbf{Naive IFC}} & \multirow[t]{4}{=}{\textbf{Sandboxing policy}} & \textit{File read/write allowlist} & All paths (\cfgval{*}) \\
   & & \textit{File read/write blocklist} & None \\
   & & \textit{Network allowlist} & All origins (\cfgval{*}) \\
   & & \textit{Network blocklist} & None \\
   & \textbf{IFC label} & \textit{Shell output} & Always \lbl{\loInt}{\hiConf} \\
   & \textbf{IFC policy} & \textit{Shell calls} & Always require approval \\
  \midrule
  \multirow[t]{11}{=}{\textbf{\ToolName}} & \multirow[t]{4}{=}{\textbf{Sandboxing policy}} & \textit{File read/write allowlist} & All paths (\cfgval{*}) \\
   & & \textit{File read/write blocklist} & None \\
   & & \textit{Network allowlist} & All origins (\cfgval{*}) \\
   & & \textit{Network blocklist} & None \\
   & \multirow[t]{5}{=}{\textbf{IFC label}} & \textit{Label functions} & \cfgval{flowsh\_default\_label.py} \\
   & & \textit{Default file label} & \lbl{\loInt}{$\{\user\}$} \\
   & & \textit{Task file labels} & \ent{\cfgval{\$CUR\_DIR}}{\lbl{\hiInt}{\loConf}}\newline
     Reproduce a GitHub issue from Teams and report the result to Teams:\newline \ent{\cfgval{\$CUR\_DIR}}{\lbl{\hiInt}{\{\user, alice-wong, bob-martinez\}}}\newline
     Implement a feature from Teams and open a pull request:\newline \ent{\cfgval{\$CUR\_DIR}}{\lbl{\hiInt}{\{\user, engineering\}}}\newline
     Implement a feature from Teams, \ldots, and report the result to Teams:\newline \ent{\cfgval{\$CUR\_DIR/credential-helper}}{\lbl{\hiInt}{\{\user, engineering\}}}\newline \ent{\cfgval{\$CUR\_DIR/aurora-service}}{\lbl{\hiInt}{\{\user, engineering, alice-wong, bob-martinez\}}}\newline
     Find related work and compare it with a private project:\newline \ent{\cfgval{\$CUR\_DIR/docs}, \cfgval{\$CUR\_DIR/experiments}}{\lbl{\hiInt}{\{\user, research\}}}\newline
     Fix a GitHub issue and open a pull request:\newline \ent{\cfgval{\$CUR\_DIR}}{\lbl{\hiInt}{\{\user, engineering\}}}\newline
     Check the references of a private paper with a tool from GitHub:\newline \ent{\cfgval{\$CUR\_DIR/workspace/paper.md}}{\lbl{\hiInt}{\{\user, research\}}} \\
   & & \textit{Network labels} & \ent{\cfgval{*}}{\lbl{\loInt}{\loConf}}\newline \ent{Package repositories}{\lbl{\hiInt}{\loConf}} \\
   & & \textit{Agent context label} & \lbl{\hiInt}{\loConf} \\
   & \multirow[t]{2}{=}{\textbf{IFC policy}} & \textit{Safe file writes} & \ent{\cfgval{*}}{\lbl{\hiInt}{\hiConf}}\newline \ent{\cfgval{\$CUR\_DIR}, \cfgval{\$TMP\_DIR}}{\lbl{\loInt}{\hiConf}}\newline \ent{\cfgblocked}{\lbl{\hiInt}{\hiConf}} \\
   & & \textit{Safe network requests} & \ent{\cfgval{*}}{\lbl{\loInt}{\loConf}, \lbl{\hiInt}{\hiConf}} \\
\end{longtable}
\endgroup
\twocolumn

\section{Defense Configurations in the Evaluation}
\label{sec:appendix-defenses}
\label{sec:appendix-configuration}

This section describes the configuration of each defense in our evaluation
(\autoref{sec:evaluation}).
\autoref{tab:defenses} lists the sandboxing policy of each defense and
the IFC labels and IFC policies of Naive IFC and \ToolName.

\PP{Lax sandbox}
The Lax sandbox allows all file reads, file writes, and network requests.
In the security-aligned setting, the evaluation adds approval requests
before every Bash call, including read-only calls, and before file writes by the agent harness's native tools
(\eg, \texttt{write\_file}) outside the working directory and temporary directory.
In the autonomy-aligned setting, the agent runs Bash calls and file writes
without approval requests.

\PP{Standard sandbox}
The Standard sandbox follows two Claude Code configurations.
Its write allowlist is the working directory and the temporary directory,
and its write blocklist is the protected paths of the Claude Code sandboxed
Bash tool~\cite{anthropicSandboxing}, such as
\texttt{\$CUR\_DIR/.git/hooks/pre-commit}, \texttt{\$CUR\_DIR/.vscode/tasks.json},
and \texttt{\$CUR\_DIR/.bashrc}.
Its network allowlist is the default allowed domains of Claude Code cloud
environments~\cite{anthropicCloudEnvironments}, which include
source-code hosting, package and container registries, cloud APIs, and
development tools.
In the security-aligned setting, the evaluation adds the same approval
requests as for the Lax sandbox, because the Standard sandbox prevents overwriting critical files but
does not prevent leaking sensitive data through the network.
In the autonomy-aligned setting, the agent runs Bash calls and file writes
without approval requests, within the sandbox policy.

\PP{Strict sandbox}
The Strict sandbox permits reads but blocks all filesystem writes, including writes
to temporary directories.
Its network allowlist permits access to package repositories.
Because its sandbox policy is restrictive enough to prevent the indirect
prompt injection attacks in our evaluation, the evaluation adds
no approval requests to the Strict sandbox in either setting.

\PP{Naive IFC}
Naive IFC allows all file reads, file writes, and network requests.
We configure it to conservatively label every shell output as \lbl{\loInt}{\hiConf}.
It requests approval before every Bash call, regardless of the
agent context label.
It requests approval for file writes by native tools outside the
configured write allowlist when the agent context label is \loInt.
In the autonomy-aligned setting, Naive IFC blocks Bash calls and file
writes that require approval.

\PP{\ToolName}
We configured the sandboxing policy of \ToolName to allow all file reads,
file writes, and network requests, so that only its IFC policy restricts them.
We configured \ToolName with the default label functions
(\autoref{sec:appendix-label-functions}).
Before each task runs, the evaluation stores the task file labels in
\autoref{tab:defenses} on the corresponding files.
In the knowledge tasks and Terminal-Bench, other files have the default file label
\lbl{\loInt}{$\{\user\}$}.
We labeled network responses as \lbl{\loInt}{\loConf},
except for responses from package repositories, which are \lbl{\hiInt}{\loConf}.

We configured the IFC policy of \ToolName to implement the two example
policies in \autoref{sec:design}.
To prevent file corruption (\autoref{sec:file-output-monitor}), the
policy defines a file write as safe if the shell context label is \hiInt,
or if the write is in the working directory (\texttt{\$CUR\_DIR}) or
temporary directory (\texttt{\$TMP\_DIR}) and not to security-sensitive files
that the Claude Code sandbox profile protects.
To prevent data exfiltration (\autoref{sec:network-output-monitor}), the
policy defines a network request to any origin as safe only if the shell
context label is \hiInt or \loConf.

In the security-aligned setting, \ToolName requests approval for file
writes and network requests that the IFC policy does not define as safe.
In the autonomy-aligned setting, \ToolName blocks them.

\newpage

\noindent\begin{minipage}{\columnwidth}
  \begin{Verbatim}[commandchars=\\\{\},codes={\catcode`\$=3\catcode`\^=7\catcode`\_=8\relax}]
\PY{k+kn}{from}\PY{+w}{ }\PY{n+nn}{collections}\PY{n+nn}{.}\PY{n+nn}{abc}\PY{+w}{ }\PY{k+kn}{import} \PY{n}{Iterable}
\PY{k+kn}{from}\PY{+w}{ }\PY{n+nn}{os}\PY{+w}{ }\PY{k+kn}{import} \PY{n}{PathLike}
\PY{k+kn}{from}\PY{+w}{ }\PY{n+nn}{typing}\PY{+w}{ }\PY{k+kn}{import} \PY{n}{Literal}

\PY{k}{def}\PY{+w}{ }\PY{n+nf}{load\PYZus{}ifc\PYZus{}policy}\PY{p}{(}\PY{n}{path}\PY{p}{:} \PY{n+nb}{str} \PY{o}{|} \PY{n}{PathLike}\PY{p}{[}\PY{n+nb}{str}\PY{p}{]}\PY{p}{)} \PY{o}{\PYZhy{}}\PY{o}{\PYZgt{}} \PY{n}{FlowshConfig}\PY{p}{:}
\PY{+w}{    }\PY{l+s+sd}{\PYZdq{}\PYZdq{}\PYZdq{}Load the configuration.\PYZdq{}\PYZdq{}\PYZdq{}}
    \PY{o}{.}\PY{o}{.}\PY{o}{.}

\PY{k}{def}\PY{+w}{ }\PY{n+nf}{load\PYZus{}file\PYZus{}label}\PY{p}{(}
    \PY{n}{path}\PY{p}{:} \PY{n+nb}{str} \PY{o}{|} \PY{n}{PathLike}\PY{p}{[}\PY{n+nb}{str}\PY{p}{]}\PY{p}{,}
    \PY{o}{*}\PY{p}{,}
    \PY{n}{namespace}\PY{p}{:} \PY{n+nb}{bool} \PY{o}{=} \PY{k+kc}{False}\PY{p}{,}
    \PY{n}{default}\PY{p}{:} \PY{n}{Label} \PY{o}{|} \PY{k+kc}{None} \PY{o}{=} \PY{k+kc}{None}\PY{p}{,}
\PY{p}{)} \PY{o}{\PYZhy{}}\PY{o}{\PYZgt{}} \PY{n}{Label} \PY{o}{|} \PY{k+kc}{None}\PY{p}{:}
\PY{+w}{    }\PY{l+s+sd}{\PYZdq{}\PYZdq{}\PYZdq{}Read a stored label or use the default.\PYZdq{}\PYZdq{}\PYZdq{}}
    \PY{o}{.}\PY{o}{.}\PY{o}{.}

\PY{k}{def}\PY{+w}{ }\PY{n+nf}{store\PYZus{}file\PYZus{}label}\PY{p}{(}
    \PY{n}{path}\PY{p}{:} \PY{n+nb}{str} \PY{o}{|} \PY{n}{PathLike}\PY{p}{[}\PY{n+nb}{str}\PY{p}{]}\PY{p}{,}
    \PY{n}{label}\PY{p}{:} \PY{n}{Label}\PY{p}{,}
    \PY{o}{*}\PY{p}{,} \PY{n}{namespace}\PY{p}{:} \PY{n+nb}{bool} \PY{o}{=} \PY{k+kc}{False}\PY{p}{,}
\PY{p}{)} \PY{o}{\PYZhy{}}\PY{o}{\PYZgt{}} \PY{k+kc}{None}\PY{p}{:}
\PY{+w}{    }\PY{l+s+sd}{\PYZdq{}\PYZdq{}\PYZdq{}Store a file label.\PYZdq{}\PYZdq{}\PYZdq{}}
    \PY{o}{.}\PY{o}{.}\PY{o}{.}

\PY{k}{def}\PY{+w}{ }\PY{n+nf}{get\PYZus{}network\PYZus{}label}\PY{p}{(}
    \PY{n}{endpoint}\PY{p}{:} \PY{n+nb}{str} \PY{o}{|} \PY{n}{CanonicalEndpoint}\PY{p}{,}
    \PY{o}{*}\PY{p}{,} \PY{n}{policy}\PY{p}{:} \PY{n}{FlowshConfig}\PY{p}{,}
\PY{p}{)} \PY{o}{\PYZhy{}}\PY{o}{\PYZgt{}} \PY{n}{Label}\PY{p}{:}
\PY{+w}{    }\PY{l+s+sd}{\PYZdq{}\PYZdq{}\PYZdq{}Look up a network input label.\PYZdq{}\PYZdq{}\PYZdq{}}
    \PY{o}{.}\PY{o}{.}\PY{o}{.}

\PY{k}{def}\PY{+w}{ }\PY{n+nf}{join\PYZus{}labels}\PY{p}{(}\PY{n}{labels}\PY{p}{:} \PY{n}{Iterable}\PY{p}{[}\PY{n}{Label}\PY{p}{]}\PY{p}{)} \PY{o}{\PYZhy{}}\PY{o}{\PYZgt{}} \PY{n}{Label}\PY{p}{:}
\PY{+w}{    }\PY{l+s+sd}{\PYZdq{}\PYZdq{}\PYZdq{}Join input labels.\PYZdq{}\PYZdq{}\PYZdq{}}
    \PY{o}{.}\PY{o}{.}\PY{o}{.}

\PY{k}{def}\PY{+w}{ }\PY{n+nf}{evaluate\PYZus{}ifc\PYZus{}policy}\PY{p}{(}
    \PY{n}{operation}\PY{p}{:} \PY{n}{Literal}\PY{p}{[}\PY{l+s+s2}{\PYZdq{}}\PY{l+s+s2}{read}\PY{l+s+s2}{\PYZdq{}}\PY{p}{,} \PY{l+s+s2}{\PYZdq{}}\PY{l+s+s2}{write}\PY{l+s+s2}{\PYZdq{}}\PY{p}{,} \PY{l+s+s2}{\PYZdq{}}\PY{l+s+s2}{network}\PY{l+s+s2}{\PYZdq{}}\PY{p}{]}\PY{p}{,}
    \PY{n}{target}\PY{p}{:} \PY{n+nb}{str} \PY{o}{|} \PY{n}{CanonicalEndpoint}\PY{p}{,}
    \PY{n}{label}\PY{p}{:} \PY{n}{Label}\PY{p}{,}
    \PY{o}{*}\PY{p}{,}
    \PY{n}{policy}\PY{p}{:} \PY{n}{FlowshConfig}\PY{p}{,}
    \PY{n}{workdir}\PY{p}{:} \PY{n+nb}{str}\PY{p}{,}
    \PY{n}{tempdir}\PY{p}{:} \PY{n+nb}{str}\PY{p}{,}
\PY{p}{)} \PY{o}{\PYZhy{}}\PY{o}{\PYZgt{}} \PY{n+nb}{bool}\PY{p}{:}
\PY{+w}{    }\PY{l+s+sd}{\PYZdq{}\PYZdq{}\PYZdq{}Check sandboxing policy and IFC policy.\PYZdq{}\PYZdq{}\PYZdq{}}
    \PY{o}{.}\PY{o}{.}\PY{o}{.}
\end{Verbatim}

  \captionof{figure}{\texttt{libflowsh} interfaces. Function bodies and some optional arguments are omitted.}
  \label{fig:flowsh-library}
\end{minipage}
\vspace{\textfloatsep}

\begin{figure}[!t]
  \input{code/flowsh_default_label}
  \caption{\texttt{flowsh\_default\_label.py} defining the default
  (Integrity, Confidentiality) labels.}
  \label{fig:default-label-functions}
\end{figure}

\section{\ToolName Library}
\label{sec:appendix-library}

The Python library \texttt{libflowsh} allows agent harnesses to apply
\ToolName's IFC labels and IFC policy to native tools.
\autoref{fig:flowsh-library} summarizes the interface.

\texttt{load\_ifc\_policy} returns a \texttt{FlowshConfig} containing the
sandboxing policy, IFC labels, and IFC policy
(\autoref{tab:flowsh-configuration}).
The agent harness passes this object as the \texttt{policy} argument to
\texttt{get\_network\_label} and \texttt{evaluate\_ifc\_policy}.
By default, a \texttt{Label} contains
\texttt{trusted} (Boolean) and \texttt{readers}
(a set of reader identifiers, or \texttt{None} for Public).
\texttt{evaluate\_ifc\_policy} checks a file read, file write, or
network request against the sandboxing policy and IFC policy.
Its target is a file path or a network origin.

\texttt{load\_file\_label} and \texttt{store\_file\_label} read and store the labels for file contents and for
filenames in the parent directory (\autoref{sec:labels}).
For files without a stored label, the agent harness passes the configured
default file label through \texttt{default}
(\autoref{tab:flowsh-configuration}).

For example, an agent harness can store a large tool result in a file and
use \texttt{store\_file\_label} to preserve its label.
When a native tool reads the file, it can use \texttt{load\_file\_label}
to obtain the file label.

\section{Default Label Functions}
\label{sec:appendix-label-functions}

\autoref{fig:default-label-functions} shows
\texttt{flowsh\_default\_label.py}, which expresses the default
(Integrity, Confidentiality) label operations from
\autoref{sec:design-policies} through the configurable Python interface.
Labels are represented as two-element JSON arrays.
Integrity is \texttt{Trusted} or \texttt{Untrusted}, and confidentiality
is \texttt{Public} or a list of permitted readers.
The label functions add \texttt{User} when interpreting a non-Public
reader list. An empty input list therefore denotes user-only data.
The top label is \loInt and readable only by \texttt{User}.

The configuration specifies the path of this file in
\texttt{label\_functions} (\autoref{tab:flowsh-configuration}).
For example, \texttt{\{"value": ["Trusted", "Public"]\}} specifies a
default file label or a policy label in the IFC policy.
\ToolName uses \texttt{leq} to check $\ell_{\mathrm{sh}} \sqsubseteq \ell$
for each policy label $\ell$ in \texttt{safe\_file\_writes} and
\texttt{safe\_network\_requests}.

\end{document}